\def\IconsInText{0}
\def\arXivIcons{0}

\def\EDIT{0}
\newcommand{\ifEd}[1]{\ifnum\EDIT=1{#1}\fi}
\newcommand{\npIf}[0]{\ifnum\EDIT=1 \newpage \fi}

\def\LuaTex{0}

\documentclass[runningheads,anonymous]{llncs}

\usepackage[T1]{fontenc}

\ifnum\EDIT=1
  \usepackage[colorinlistoftodos]{todonotes}
\else
  \usepackage[colorinlistoftodos,disable]{todonotes}
\fi

\setuptodonotes{fancyline}

\newcommand{\ifepr}[0]{\ifnum\EPRINT=1}
\newcommand{\ififs}[0]{\ifnum\IFIPSEC=1}

\ifnum\IconsInText=0
  \newcommand{\qoesign}[0]{QoeSiGN\xspace}
\else
  \newcommand{\qoesign}[0]{$\mathbb{Q}${\oe}SiGN\xspace}
\fi

\newcommand{\circEmpty}[1]{\tikz[baseline=-0.75ex] \draw[draw=#1, fill=gray!20, thick] (0,0) circle (.42em);}
\newcommand{\circEmptyTab}[1]{\tikz[baseline=-1.00ex] \draw[draw=#1, fill=gray!20, thick] (0,0) circle (.42em);}

\newcommand{\circSmallDot}[1]{%
  \tikz[baseline=-0.75ex]{
    \draw[draw=#1, fill=#1!20, thick] (0,0) circle (.42em);
    \fill[#1] (0,0) circle (.16em);
  }%
}

\newcommand{\circCrossRed}[2]{
  \tikz[baseline=#2ex] {
    \draw[draw=#1, fill=red!20, thick] (0,0) circle (.42em);
    \draw[#1, thick] (-0.18em,-0.18em) -- (0.18em,0.18em);
    \draw[#1, thick] (-0.18em,0.18em) -- (0.18em,-0.18em);
    }
}

\newcommand{\circFull}[1]{\tikz[baseline=-0.75ex] \draw[draw=#1, fill=#1, thick] (0,0) circle (.42em);}
\newcommand{\circFullTab}[1]{\tikz[baseline=-1.00ex] \draw[draw=#1, fill=#1, thick] (0,0) circle (.42em);}

\newcommand{\circHalfBot}[1]{\circHalf{#1}{-0.75ex}}
\newcommand{\circHalfBotTab}[1]{\circHalf{#1}{-1.00ex}}
\newcommand{\circHalf}[2]{
  \tikz[baseline=#2]{
    \fill[#1] (.42em,0) arc[start angle=360, end angle=180, radius=.42em] -- cycle;
    \draw[draw=#1, thick] (0,0) circle(.42em);}\xspace}

\newcommand{\npIfEdit}[0]{\ifnum\EDIT=1 \newpage \fi}

\newcommand{\paragraphBold}[1]{\paragraph{\textnormal{\textbf{#1.}}}}
\newcommand{\parB}[1]{\paragraphBold{#1}}

\newcommand{\paragraphBoldExt}[1]{
  \ifnum\extendedText=1
    \paragraph{\textnormal{\textbf{\ext{#1.}}}}
  \else\fi
}

\newcommand{\fn}[1]{\footnote{#1}}
\newcommand{\lbr}[0]{\linebreak}

\newcommand{\mc}[3]{\multicolumn{#1}{#2}{#3}}
\newcommand{\mr}[3]{\multirow{#1}{#2}{#3}}

\newcommand{\cFull}[0]{\cellGreen \vspace{.15em}\circFull{circleGreen}}
\newcommand{\cYes}[0]{\cellGreen \vspace{.15em}\circFull{circleGreen}}
\newcommand{\cHalf}[0]{\cellBlue \vspace{.15em}\circHalfBot{enumBlue}}
\newcommand{\cNo}[0]{\cellGray \vspace{.15em}\circEmpty{circleGray}}
\newcommand{\cNoRed}[0]{\cellRed \vspace{.15em}\circCrossRed{circleRed}{-0.75}}

\newcommand{\cellBlue}[0]{\cellcolor{enumBlue!25}}
\newcommand{\cellGreen}[0]{\cellcolor{green!20}}
\newcommand{\cellGray}[0]{\cellcolor{gray!20}}
\newcommand{\cellOrange}[0]{\cellcolor{threatPrioHigh!75}}

\newcommand{\cellRed}[0]{\cellcolor{red!20}}
\newcommand{\cellYellow}[0]{\cellcolor{tableAttentionYellow!50}}

\newcommand{\rarr}[0]{$\rightarrow$\xspace}

\newcommand{\Rarr}[0]{$\Rightarrow$\xspace}
\newcommand{\RarrB}[0]{$\boldsymbol{\Rightarrow}$\xspace}

\newcommand{\hrefU}[2]{\href{#1}{\underline{#2}}}
\newcommand{\nbrSp}{\nobreakspace}
\newcommand{\quoteInline}[1]{\textit{\enquote{#1}}}

\newcommand{\ttt}[1]{\texttt{#1}}
\newcommand{\tbf}[1]{\textbf{#1}}
\newcommand{\tit}[1]{\textit{#1}}

\newcommand{\mystrut}{\rule[-1.75pt]{0pt}{9.25pt}}
\newcommand{\cboxRqMust}[1]{
  \ifnum\IconsInText=1
  \setlength{\fboxsep}{2pt}\colorbox{rqMust}{\mystrut\color{white}#1}\xspace
  \else
    #1
  \fi
}
\newcommand{\cboxRqBacklog}[1]{\ifnum\IconsInText=1\setlength{\fboxsep}{2pt}\colorbox{rqBacklog}{\mystrut#1}\xspace\else#1\fi}
\newcommand{\cboxMIT}[1]{
  \ifnum\IconsInText=1
    \setlength{\fboxsep}{2pt}\colorbox{blue!20}{\mystrut#1}\xspace
  \else
  #1
  \fi
}
\newcommand{\mystrutPC}{\rule[-1.5pt]{0pt}{9pt}}
\newcommand{\cboxPC}[1]{\ifnum\IconsInText=1\setlength{\fboxsep}{1pt}\colorbox{p2c2Blue}{\mystrutPC\color{white}#1}\xspace\hspace{-0.5em}\else#1\fi}

\newcommand{\td}[1]{\todo{#1}}
\newcommand{\tdi}[1]{\todo[inline]{#1}}

\newcommand{\tbd}[0]{\tdi{tbd:)}}

\newcommand{\todoInf}[1]{\todo[color=grassGreen]{\textcolor{white}{INFO: #1}}}
\newcommand{\tdInf}[1]{\todoInf{#1}}

\newcommand{\todoFin}[1]{\todo[color=orange!60]{\textcolor{white}{FIN: #1}}}

\newcommand{\tdFin}[1]{\todoFin{#1}}

\usepackage{footmisc}

\usepackage{booktabs}

\usepackage{adjustbox}

\usepackage{hhline}

\usepackage{array}

\usepackage{multirow}

\usepackage{makecell}

\usepackage{tabularx}
\newcommand{\tn}[0]{\tabularnewline}

\usepackage[inline]{enumitem}

\usepackage{amsmath}
\usepackage{amsfonts}
\usepackage{amssymb}

\ifnum\LuaTex=1
\usepackage{fontspec}
\newfontfamily\titleFont{Libertinus Serif}

\fi

\usepackage{tikzsymbols}
\usepackage{fontawesome5} 
\usepackage{pifont} 

\usepackage{xpatch}
\makeatletter
\xpatchcmd{\@cref}{\begingroup}{\begingroup\color{htmlBlue}}{}{}
\makeatother

\usepackage[table]{xcolor}

\definecolor{codeblue}{HTML}{21618C}
\definecolor{enumBlue}{HTML}{2E86C1}
\definecolor{htmlBlue}{RGB}{52,152,219}
\definecolor{indigoBlue}{HTML}{1560BD} 
\definecolor{skyBlue}{RGB}{174,214,241}
\definecolor{indigoLikeBlue}{HTML}{1571ED}
\definecolor{ePrintBlue}{HTML}{0070F0}
\definecolor{p2c2Blue}{HTML}{5DADE2}

\definecolor{circleGray}{HTML}{797986} 
\definecolor{codegray}{rgb}{0.5,0.5 0.5}
\definecolor{gray}{rgb}{0.5,0.5,0.5}
\definecolor{silver}{RGB}{192,192,192}

\definecolor{codegreen}{rgb} {0,0.6,0}
\definecolor{dkgreen}{rgb}{0,0.6,0}
\definecolor{grassGreen}{RGB}{218,247,166}
\definecolor{tabBackGreen}{RGB}{152, 251, 152}

\definecolor{circleGreen}{RGB}{49, 201, 80} 

\definecolor{CANextendedColor}{HTML}{A569BD}
\definecolor{CANSorange}{HTML}{EB984E}

\definecolor{codepurple}{rgb}{0.58,0,0.82}
\definecolor{mauve}{rgb}{0.58,0,0.82}

\definecolor{circleRed}{HTML}{FF8A8C}

\definecolor{shortTurquoise}{HTML}{45B39D}

\definecolor{backcolour}{rgb}{0.95,0.95,0.92}
\definecolor{code_backcolor}{rgb}{0.95,0.95,0.92}

\definecolor{tableAttentionYellow}{HTML}{F6D965}

\definecolor{threatPrioHigh}{HTML}{E97132} 
\definecolor{threatPrioMedium}{HTML}{FFD230} 
\definecolor{threatPrioLow}{HTML}{83E28E} 

\definecolor{rqMust}{HTML}{9810FA} 
\definecolor{colorRqMust}{HTML}{F6D965} 
\definecolor{rqAware}{HTML}{CE8CFD} 
\definecolor{rqBacklog}{HTML}{C0E6F5} 

\usepackage{xurl}
\usepackage[breaklinks]{hyperref}
\hypersetup{
  colorlinks=true,
  urlcolor=enumBlue, 
  linkcolor=enumBlue 
}

\usepackage[nameinlink]{cleveref}

\makeatletter
\AtBeginDocument
 {
   \def\ltx@label#1{\cref@label{#1}}
   \def\label@in@display@noarg#1{\cref@old@label@in@display{#1}}
\def\label@in@mmeasure@noarg#1{%
    \begingroup%
      \measuring@false%
      \cref@old@label@in@display{#1}
    \endgroup}%
 } %
\makeatother

\makeatletter
\AddToHook{cmd/appendix/before}{\crefalias{section}{appendix}}
\makeatother

\usepackage[htt]{hyphenat}

\usepackage{rotating}

\usepackage{graphicx}

\usepackage{subcaption}

\usepackage{listings}
\lstdefinestyle{dot-product}{
    backgroundcolor=\color{code_backcolor},
    commentstyle=\color{codegreen},
    keywordstyle=\color{magenta},
    numberstyle=\tiny\color{codegray},
    stringstyle=\color{dkgreen},
    basicstyle=\ttfamily\footnotesize,
    breakatwhitespace=false,
    captionpos=b,
    keepspaces=true,
    numbers=left,
    numbersep=5pt,
    showspaces=false,
    showstringspaces=false,
    showtabs=false,
    tabsize=2,
    morecomment=[s][{\color{enumBlue}}]{<}{>},
    morecomment=[s][{\color{enumBlue}}]{</}{>},
    morecomment=[s][{\color{enumBlue}}]{__}{__},
    morecomment=[s][{\color{magenta}}]{@for_}{range}, 
    morecomment=[s][{\color{magenta}}]{sin}{t}, 
    morecomment=[s][{\color{enumBlue}}]{start_t}{)},
    morecomment=[s][{\color{enumBlue}}]{stop_t}{)},
}

\lstdefinestyle{b4Mformat}{
    backgroundcolor=\color{backcolour},
    commentstyle=\color{codegreen},
    keywordstyle=\color{magenta},
    numberstyle=\tiny\color{codegray},
    stringstyle=\color{dkgreen},
    basicstyle=\ttfamily\footnotesize,
    breakatwhitespace=false,
    captionpos=b,
    keepspaces=true,
    numbers=left,
    numbersep=5pt,
    showspaces=false,
    showstringspaces=false,
    showtabs=false,
    tabsize=2,
    morecomment=[s][{\color{enumBlue}}]{<}{>},
    morecomment=[s][{\color{enumBlue}}]{</}{>},
    morecomment=[s][{\color{magenta}}]{@for_}{range}, 
    morecomment=[s][{\color{dkgreen}}]{start_t}{imer},
    morecomment=[s][{\color{dkgreen}}]{stop_t}{imer},
}

\lstdefinestyle{b4Mcommands}{
    backgroundcolor=\color{backcolour},
    commentstyle=\color{codegreen},
    keywordstyle=\color{magenta},
    numberstyle=\tiny\color{codegray},
    stringstyle=\color{dkgreen},
    basicstyle=\ttfamily\footnotesize,
    breakatwhitespace=false,
    captionpos=b,
    keepspaces=true,
    numbers=left,
    numbersep=5pt,
    showspaces=false,
    showstringspaces=false,
    showtabs=false,
    tabsize=2,
    morecomment=[s][{\color{enumBlue}}]{<}{>},
    morecomment=[s][{\color{enumBlue}}]{</}{>},
    morecomment=[s][{\color{enumBlue}}]{(}{)},
    morecomment=[s][{\color{enumBlue}}]{__}{__},
    morecomment=[s][{\color{magenta}}]{@for_}{range}, 
    morecomment=[s][{\color{dkgreen}}]{start_t}{imer},
    morecomment=[s][{\color{dkgreen}}]{stop_t}{imer},
}

\usepackage{csquotes}

\usepackage{setspace}

\usepackage{moreverb}

\usepackage{mdframed}

\usepackage{svg}

\AtEndPreamble{%
  \RequirePackage{hyperref}
  \hypersetup{%
        colorlinks    = true,
        citecolor     = codegreen,
        urlcolor      = enumBlue 
      }
}

\usepackage{xspace}

\usepackage{soul} 
\sethlcolor{code_backcolor}
\makeatletter
 \def\SOUL@hlpreamble{%
 \setul{}{2.5ex}
 \let\SOUL@stcolor\SOUL@hlcolor
 \SOUL@stpreamble
 }
\makeatother

\newcommand{\lnRef}[1]{{\Cref{#1} \textit{(\nameref{#1})}}}

\newcommand{\lbrk}[0]{\linebreak\xspace}

\newcommand{\relaxTT}[1]{{\let\ttt\relax#1}}

\usepackage{acronym}

\newacro{mpc}[MPC]{Secure Multi-Party Computation}

\newacro{fhe}[FHE]{Fully Homomorphic Encryption}

\newacro{he}[HE]{Homomorphic Encryption}

\newacro{pet}[PET]{privacy-enhancing technologies}

\newacro{tee}[TEE]{trusted execution environment}

\newcommand{\teeP}{\acp{tee}\xspace}

\newacro{zkp}[ZKP]{zero-knowledge proof}

\newacro{b4m} [\ttt{b4M}] {\ttt{Holistic} \ttt{Benchmarking} \ttt{for} \ttt{MPC}}

\newacro{scale} [\ttt{SCALE}] {\ttt{Secure Computation Algorithms from LEuven}}
\newacro{scamba} [\ttt{SCALE-MAMBA}] {{\ac{scale}\ttt{--Multiparty AlgorithMs Basic Argot}} \acused{scale}}
\newacro{scrust} [\ttt{SCALE-Rust}] {{\ac{scale}\ttt{--Rust}} \acused{scale}}
\acused{scale-rust}
\newacro{mp-spdz} [\ttt{MP-SPDZ}] {{\ttt{Multi-Protocol SPDZ}}}
\newacro{spdz-2} [\ttt{SPDZ-2}] {{\ttt{SPDZ-2}}}
\newacro{fresco} [\ttt{FRESCO}] {{\ttt{FRamework for Efficient and Secure COmputation}}}
\newacro{mpyc} [\ttt{MPyC}] {\ttt{Multiparty Computation in Python}}
\newacro{motion} [\ttt{MOTION}] {\ttt{Framework for Mixed-Protocol Multi-Party Computation}}

\newacro{prf}[PRF]{pseudo-random function\acused{prf_attr}}
\newacro{prf_attr}[PRF]{pseudo-random-function\acused{prf}}
\newacro{ppml}[PPML]{privacy-preserving ML}

\newacro{mpspdz}[\ttt{MP-SPDZ}]{``Multi-Protocol SPDZ - A Versatile Framework for MPC''}
\newcommand{\mpsz}{\ac{mpspdz}\xspace}
\newacro{nSZ}[\ttt{natSPDZ}]{``Native Variant of Multi-Protocol-SPDZ''}

\newacro{wSZ}[\ttt{webSPDZ}]{``Web Variant of Multi-Protocol-SPDZ''}
\newcommand{\wsz}{\ac{wSZ}\xspace}

\newacro{mpyc}[\ttt{MPyC}]{``Multi-Party Computation in Python''}
\newcommand{\mpyc}{\ac{mpyc}\xspace}
\newacro{mpyN}[\ttt{MPyC-Nat}]{``Native Variant of Multi-Party Computation in Python''}

\newacro{mpyW}[\ttt{MPyC-Web}]{``Web Variant of Multi-Party Computation in Python''}
\newcommand{\mpyw}{\ac{mpyW}\xspace}

\newacro{JF}[\ttt{JIFF}]{``JavaScript library for building web apps that employ MPC''}

\newacro{emsc}[\ttt{Emscripten}]{\ttt{Emscripten}}

\newacro{logreg}[LogReg]{logistic regression}

\newacro{browser}[web browser]{web browser}

\newacro{wasm}[Wasm]{WebAssembly}

\newacro{simd}[SIMD]{``Single Instruction, Multiple Data''}

\newacro{wrtc}[\ttt{WebRTC}]{``Web Real-Time Communication''}

\newacro{srtp}[SRTP]{Secure Real-time Transport Protocol \acused{srtpThe}}
\newacro{srtpThe}[SRTP]{the Secure Real-time Transport Protocol \acused{srtp}}
\newacro{p2p}[P2P]{peer to peer\acused{p2pH}}
\newacro{p2pH}[P2P]{peer-to-peer\acused{p2p}}

\newacro{eid}[eID]{electronic identity}
\newacroplural{eid}[eIDs]{electronic identities}
\newcommand{\eid}{\ac{eid}\xspace}

\newcommand{\ida}{ID Austria\xspace}

\newacro{eidas}[eIDAS]{\enquote{electronic IDentification, Authentication and trust Services}}
\newcommand{\eidas}{\ac{eidas}\xspace}

\newacro{eudiw}[EUDI wallet]{EU Digital Identity Wallet}

\newacro{hsm}[HSM]{hardware security module}
\newcommand{\hsm}{\ac{hsm}\xspace}
\newcommand{\hsmP}{\acp{hsm}\xspace}

\newacro{qscd}[QSCD]{Qualified Signature Creation Device}
\newcommand{\qscd}{\ac{qscd}\xspace}

\newacro{qtsp}[QTSP]{Qualified Trust Service Provider}
\newcommand{\qtsp}{\ac{qtsp}\xspace}
\newcommand{\qtspP}{\acp{qtsp}\xspace}

\newacro{sigSes}[SES]{\emph{Simple Electronic Signature}}

\newacro{sigAes}[AdES]{\emph{Advanced Electronic Signature}}

\newacro{sigQes}[QES]{\emph{Qualified Electronic Signature}}

\newacro{sp}[SP]{Service Provider}

\newcommand{\sepP}{\acp{sp}\xspace}

\newacro{sso}[SSO]{single sign-on}

\newacro{owasp}[OWASP]{Open Web Application Security Project}
\newcommand{\owasp}{\ac{owasp}\xspace}

\newacro{dfd}[DFD]{Data Flow Diagram}
\newcommand{\dfd}{\ac{dfd}\xspace}

\newcommand{\dreadLink}{\hrefU{https://en.wikipedia.org/wiki/DREAD_(risk_assessment_model)}{DREAD}\xspace}

\newcommand{\stride}{STRIDE\xspace}
\newcommand{\strideLink}{\hrefU{https://en.wikipedia.org/wiki/STRIDE_model}{STRIDE}\xspace}

\usepackage{textcomp}

\newcommand{\ti}[1]{\textit{#1}}

\usepackage{tcolorbox}

\usepackage{orcidlink}

\usepackage{float}

\newcommand{\ext}[1]{\ifnum\extendedText=1{\ifnum\HLextended=1{\color{ePrintBlue}{#1}}\else#1\fi}\else{}\fi}

\newcommand{\extCANS}[1]{\ifnum\CANextended=1{\ifnum\hlCANextended=1{\color{CANextendedColor}{#1}}\else#1\fi}\else{}\fi}

\newcommand{\extA}[2]{\ifnum\extendedText=1{\ifnum\HLextended=1{{\color{ePrintBlue}#1}}\else#1\fi}\else{\ifnum\HLshort=1{{\color{shortTurquoise}#2}}\else#2\fi}\fi}

\newcommand{\extCANSadd}[2]{\ifnum\CANextended=1{\ifnum\hlCANextended=1{\color{CANextendedColor}{#1}}\else#1\fi}\else{\ifnum\HLshort=1{{\color{shortTurquoise}#2}}\else#2\fi}\fi}

\newcommand{\short}[1]{\ifnum\extendedText=0{\ifnum\HLshort=1{\color{shortTurquoise}{#1}}\else#1\fi}\else{}\fi}

\newcommand{\cans}[1]{\ifnum\CANS=1{\ifnum\hlCANS=1{\color{CANSorange}{#1}}\else{#1}\fi}\else{}\fi}

  \usepackage[user, lastpage]{zref}
  \usepackage{etoolbox}

  \usepackage{xassoccnt}
  \NewTotalDocumentCounter{totalpages}

  \usepackage{totpages}
  \usepackage{fancyhdr}
\ifnum\EDIT=1
  \fancypagestyle{headings}{
  }
\else
\fi

\begin{document}
\ifnum\IconsInText=0
\newcommand{\faShieldText}{}
\newcommand{\faMountainText}{}
\newcommand{\faSortAmountUpText}{}
\newcommand{\faSearchText}{}
\newcommand{\faCogsText}{}
\newcommand{\faRunningText}{}
\newcommand{\faHandsHelpingText}{}
\newcommand{\faYinYangText}{}
\newcommand{\faAtomText}{}
\newcommand{\faExchangeText}{}
\newcommand{\faKeyText}{}
\newcommand{\faUserFriendsText}{}
\newcommand{\faBookText}{}
\newcommand{\faStopwatchText}{}
\newcommand{\faBalanceScaleText}{}
\newcommand{\faFileSignatureText}{}
\newcommand{\faAwardText}{}
\newcommand{\faYinYangArXiv}{}

\renewcommand{\strideLink}{STRIDE\xspace}
\renewcommand{\dreadLink}{DREAD\xspace}
\newcommand{\threatModelProcessLink}{threat modeling process\xspace}
\newcommand{\thresholdCryptoArXiv}{threshold cryptography\xspace}
\newcommand{\mpcArXiv}{secure multi-party computation (MPC)\xspace}
\newcommand{\heArXiv}{homomorphic encryption (HE)\xspace}
\newcommand{\hsmArXiv}{\Acf{hsm}\xspace}
\newcommand{\teeArXiv}{\Acf{tee}\xspace}
\newcommand{\qessLinkArXiv}{Qualified Electronic Signatures (QESs)\xspace}
\newcommand{\adesLinkArXiv}{Advanced Electronic Signature (AdES)\xspace}
\newcommand{\kerckhoffLinkArXiv}{Kerckhoff's principle\xspace}

\newcommand{\colorizeText}[1]{}

\newcommand{\contentNonIcon}[1]{#1}
\fi

\ifnum\IconsInText=1
\ifnum\arXivIcons=0
\newcommand{\faShieldText}{\faShield*}
\newcommand{\faMountainText}{\faMountain}
\newcommand{\faSortAmountUpText}{\faSortAmountUpText}
\newcommand{\faSearchText}{\faSearch}
\newcommand{\faCogsText}{\faCogs}
\newcommand{\faRunningText}{\faRunning}
\newcommand{\faHandsHelpingText}{\faHandsHelping}
\newcommand{\faYinYangText}{\faYinYang}
\newcommand{\faAtomText}{\faAtom}
\newcommand{\faExchangeText}{\faExchange*}
\newcommand{\faKeyText}{\faKey}
\newcommand{\faUserFriendsText}{\faUserFriends}
\newcommand{\faBookText}{\faBook}
\newcommand{\faStopwatchText}{\faStopwatch}
\newcommand{\faBalanceScaleText}{\faBalanceScale}
\newcommand{\faFileSignatureText}{\faFileSignature}
\newcommand{\faAwardText}{\faAward}
\renewcommand{\faSearch}{}
\newcommand{\faYinYangArXiv}{}
\renewcommand{\strideLink}{STRIDE\xspace}
\renewcommand{\dreadLink}{DREAD\xspace}
\newcommand{\threatModelProcessLink}{threat modeling process\xspace}
\newcommand{\thresholdCryptoArXiv}{threshold cryptography\xspace}
\newcommand{\mpcArXiv}{secure multi-party computation (MPC)\xspace}
\newcommand{\heArXiv}{homomorphic encryption (HE)\xspace}
\newcommand{\hsmArXiv}{\Acf{hsm}\xspace}
\newcommand{\teeArXiv}{\Acf{tee}\xspace}
\newcommand{\qessLinkArXiv}{Qualified Electronic Signatures (QESs)\xspace}
\newcommand{\adesLinkArXiv}{Advanced Electronic Signature (AdES)\xspace}
\newcommand{\kerckhoffLinkArXiv}{Kerckhoff's principle\xspace}
\newcommand{\faYinYangArXiv}{\faYinYang\nbrSp}

\newcommand{\colorizeText}[1]{\color{#1}}
\newcommand{\contentNonIcon}[1]{}

\else 
\newcommand{\faYinYangArXiv}{\faYinYang\nbrSp}
\newcommand{\threatModelProcessLink}{\hrefU{https://owasp.org/www-community/Threat_Modeling_Processthreat}{threat modeling process\xspace}}
\newcommand{\thresholdCryptoArXiv}{\hrefU{https://en.wikipedia.org/wiki/Threshold_cryptosystem}{threshold cryptography\xspace}}
\newcommand{\mpcArXiv}{\hrefU{https://en.wikipedia.org/wiki/Secure_multi-party_computation}{secure multi-party computation (MPC)\xspace}}
\newcommand{\heArXiv}{\hrefU{https://en.wikipedia.org/wiki/Homomorphic_encryption}{homomorphic encryption (HE)\xspace}}
\newcommand{\hsmArXiv}{\hrefU{https://en.wikipedia.org/wiki/Hardware_security_module}{\Acf{hsm}\xspace}}
\newcommand{\teeArXiv}{\hrefU{https://en.wikipedia.org/wiki/Trusted_execution_environment}}{\Acf{tee}\xspace}}
\newcommand{\qessLinkArXiv}{\hrefU{https://en.wikipedia.org/wiki/Qualified_electronic_signature}{Qualified Electronic Signatures (QESs)\xspace}}
\newcommand{\adesLinkArXiv}{\hrefU{https://en.wikipedia.org/wiki/Advanced_electronic_signature}{Advanced Electronic Signature (AdES)\xspace}}
\newcommand{\kerckhoffLinkArXiv}{\hrefU{https://en.wikipedia.org/wiki/Kerckhoffs's_principle}{Kerckhoff's principle\xspace}}
\fi
\fi

\ifnum\EDIT=1
  
\setcounter{tocdepth}{3} 
\tableofcontents

\newpage
\listoffigures
\listoftables

\newpage
\begin{tcolorbox}[colback=silver, boxrule=0pt, left=5pt, right=5pt, top=5pt, bottom=5pt]
  \color{white}
  \listoftodos
  \color{black}
\end{tcolorbox}

\setcounter{page}{0}

  \newpage
\else
\fi

\ifnum\EDIT=1
  \thispagestyle{headings}
  \pagestyle{headings}
\else
\fi

\ifnum\LuaTex=0
  \ifnum\IconsInText=1
    \title{\faYinYangArXiv\nobreakspace$\mathbb{Q}${\oe}SiGN: \\Towards Qualified Collaborative eSignatures}
  \else
    \title{QoeSiGN: \\Towards Qualified Collaborative eSignatures}
  \fi
\else
    \title{\titleFont{\char"211A}{\oe}SiGN: \\Qualified Collaborative eSignatures}
\fi


%
%

\author{
  Karl W. Koch\inst{1,2}\orcidlink{0000-0002-4828-7587},\\ 
  Stephan Krenn\inst{3}\orcidlink{0000-0003-2835-9093},
  Alexandra Hofer\inst{3}
}


%
%
\institute{
  Graz University of Technology, Austria
  \and
  Secure Information Technology Center Austria (A-SIT), Graz, Austria
  \\ \email{karl.koch@tugraz.at}
  \and
  AIT Austrian Institute of Technology GmbH, Vienna, Austria
  \\ \email{stephan.krenn@ait.ac.at}
  \and
  LIT Law Lab, Johannes Kepler University Linz, Austria
  \\ \email{alexandra.hofer@jku.at}
}


\ifnum\EDIT=1
  \thispagestyle{headings}
  \pagestyle{headings}
\fi

\ifnum\EDIT=1
  \zlabel{startcount}
\fi

\maketitle              


\begin{abstract}
{
 \setlength{\parindent}{2em}\setlength{\parskip}{0pt}%

eSignatures ensure data's authenticity, non-repudiation, and integrity.
EU's eIDAS regulation specifies, e.g., 
advanced 
and qualified (QES) eSignatures. 
While eSignatures' concrete legal effects depend on the individual case, QESs constitute the highest level of technical protection and authenticity under eIDAS.
QESs are based on a qualified certificate issued by a qualified trust service provider (QTSP). 
Despite 
legal requirements, technically, a QTSP represents a single point of failure due to its sole responsibility for computing a QES. 
Contrary to a single computation entity, privacy-preserving collaborative computations (P2C2s) have become increasingly practical in recent years; yet lacking an extensive investigation on potential integrations in the QES landscape. 

\indent\tbf{This Work.}
We perform a \faSearchText threat analysis on the QES-creation process of Austria's national eID, using STRIDE 
and a DREAD-like model to extract \faMountainText requirement challenges (RCs)
primarily related to: \lbrk
{\colorizeText{rqMust} \faCogsText{}\contentNonIcon{(1)} Distributed Service Robustness, 
\faRunningText{}\contentNonIcon{(2)} Agile Crypto Deployment, 
and \faHandsHelpingText{}\contentNonIcon{(3)} Active User Involvement.} 
To address these RCs, we present \lbrk\faYinYangText\qoesign, utilizing novel P2C2 technologies.
While \emph{currently} no P2C2 addresses all RCs, legal aspects, and practical efficiency simultaneously, \faYinYangText\qoesign gives instantiation possibilities for different needs.
For instance, 
\enquote{Multi-Party HSMs}
for distributed hardware-secured computations; 
or secure multi-party computation (software) for highest crypto agility and \lbrk user involvement, where the user participates in the QES computation.

Deployment-wise, QTSPs would need to adapt the signing process and setup trusted communication channels. 
Legal-wise, \faYinYangText\qoesign's \lbrk implementation appears permissible
, needing further analysis for \lbrk realization.
Technically, \faYinYangText\qoesign addresses some regulation requirements better than the current solution, such as \enquote{sole control} or crypto agility.

\faYinYangText\qoesign is not limited to \ida.
Our identified threats and extracted requirements can be transferred to the general QES ecosystem.
\vspace{-0.25em}

}
\end{abstract}


\npIf

\section{Introduction}
\label{sec-Intro}
\acresetall

\tdInf{WHY}
\tdInf{contribution -> flexible, robust and user-centric QES (eID Services later on)}
\td{v3/4: Grammarly all the way :D}

\noindent
In an increasingly digitized world, e(lectronic)Signatures are essential to secure digital communications and transactions. 
Whether, e.g., transactions in the growing cryptocurrency domain, signing contracts in ever-expanding digitized and remote (business) processes, or enhanced auditability (who signed when); eSignatures are of ubiquitous importance to verify the data's origin (authenticity and non-repudiation) and protect against tampering or forgery (integrity). 

\paragraphBold{Electronic Signatures}
At the level of European Union (EU) law, the \eidas regulation~\cite{eIDAS}\fn{Regulation (EU) No 910/2014 and Regulation (EU) No 2024/1183\label{fn-eIDAS}} constitutes the primary legal framework for eSignatures. 
As an EU regulation, \eidas is directly applicable in all member states, meaning its provisions are binding. 
\eidas defines eSignatures as \quoteInline{data in electronic form [...] logically associated with other data [...] used by the signatory to sign}.
This \enquote{basic form} is commonly referred to as simple electronic signature (SES), consisting of, e.g., signature images or even a typed name. 
Further, \eidas defines advanced (AdES), needing to fulfill specific security requirements, and qualified eSignatures (QES), comprising additional requirements, such as being based on a qualified certificate issued by a \qtsp.
While all three levels can provide legal value, depending on the respective regulation in the application domain, QESs constitute the highest level of technical protection and authenticity under \eidas. 
\td{re-check for arXiv :D As such, QESs are required for certain \enquote{high-value} contracts, such as (Austria's) \enquote{surety} (cf. Art 1346 and 886 ABGB + Art 4 SVG).}

Moreover, \eidas establishes the principle that QESs \emph{may be} equivalent to handwritten signatures (cf. Art 25 para 2\footref{fn-eIDAS}).
While deviations from this principle may be made both by law and by contract, in general, it has been incorporated into, e.g., Austrian law (cf. Art 4 SVG).
See~\Cref{appx-QES-handwritten-equivalence} for further details on this \emph{general} equivalence principle for QESs and handwritten signatures.
\paragraphBold{Single Qualified Signing Service}
A \qtsp must be officially recognized and comply with strict regulatory requirements~\cite{eIDAS}.
For instance, it must implement strong security measures to protect private keys and guarantee the authenticity, integrity, and confidentiality of signature creation data.
As of now, trust in a \qtsp relies on its certification, regular audits, and adherence to the \eidas regulation to maintain legal validity of signatures.

However, despite strong procedural and legal requirements, technically , any \qtsp also acts as a single point of failure because of its sole responsibility for issuing and managing QESs. 
Specifically, if a \qtsp experiences technical \lbrk failures, security breaches, or becomes compromised, the availability, validity, and trustworthiness of all signatures relying on its certificates may be compromised. 
Thus, this centralization creates a critical dependency, where any disruption or compromise at the \qtsp level can undermine the entire trust framework.

\paragraphBold{Advanced Collaborative Cryptography}
In contrast to a single computation \lbrk entity, privacy-preserving collaborative computations (P2C2s) have become \lbrk increasingly practical in recent years, which enable joint computations among many entities while preserving privacy.
P2C2s are enabled by advanced cryptography, such as threshold computations or homomorphic encryption. 
Besides \lbrk\enquote{Split-ECDSA}~\cite{DBLP:journals/iacr/Verheul21_secdsa}, which focuses on user-integrated signing with a single \lbrk instance; i.e., \qtsp for (identity) wallets, we are not aware of extensive \lbrk(academic) work or studies addressing P2C2 integrations into the QES-creation process.
Further, Split-ECDSA leaves the legal implementation for user devices as future work and uses a non-Post-Quantum-secure (non-PQ) algorithm.


\subsection{Our Contributions}
\tdInf{HOW \& WHAT :)}

\noindent
This paper
overcomes the limitations caused by the single-point-of-failure of \qtspP.
We utilize cryptographic mechanisms to decentralize signature creation.

\paragraphBold{\faSearchText QES-Flow Threat Analysis and \faMountainText Requirement Challenges (RCs)}
\nbrSp\lbrk We perform a threat analysis of \ida's QES-creation process; 
\lbrk investigating essential flows, processes and data stores concerning the \qtsp's remote signature-creation device.
This analysis reveals RCs primarily related to:
{
  \ifnum\IconsInText=1
    \renewcommand{\labelitemi}{\faMountainText}
    \renewcommand{\labelenumi}{\faMountainText}
  \else
    \renewcommand{\labelenumi}{\hspace{0.5em}\textbf{(\theenumi)}}
  \fi
  \begin{enumerate}
    \item \cboxRqMust{\ifnum\IconsInText=1\tbf{(1)}\fi \faCogsText{}\textbf{Distributed Service Robustness}}\emph{(\faSortAmountUpText security, availability, and stability)}, 
    \item \cboxRqMust{\ifnum\IconsInText=1\tbf{(2)}\fi \faRunningText{}\textbf{Agile Crypto Deployment}}\emph{(\faSortAmountUpText (post-quantum) security)}, and
    \item \cboxRqMust{\ifnum\IconsInText=1\tbf{(3)}\fi \faHandsHelpingText{}\textbf{User Involvement}}\emph{(\faSortAmountUpText trust and transparency)}.
  \end{enumerate}
}
\ifnum\IconsInText=0
  \paragraphBold{\qoesign Concept}
\else
  \paragraphBold{\faYinYangText\nobreakspace\qoesign Concept}
\fi
To address our extracted RCs, we introduce \qoesign.
Our concept comes in various flavors, offering different trade-offs by utilizing privacy-preserving collaborative computation based on novel cryptography.
\lbrk Primarily, we leverage either hardware-software hybrids (e.g., \enquote{Multi-Party HSM}) or pure software-based approaches (e.g., \enquote{MPC-based Virtual Security Modules} (VSMs)).
Additionally, we investigate \qoesign's legal compatibility and performance aspects from a theoretical point of view
; followed by our conclusions. 

\section{\faSearchText Threat Analysis of ID Austria's QES Creation Flow}
\label{sec-threat-modelling}
\quoteInline{Anyone who is concerned about the privacy, safety, and security of their system} should \emph{\enquote{threat model}}, states the Threat Modeling Manifesto~\cite{threat_modeling_manifesto}.
A structured \threatModelProcessLink highlights valuable components and respective flows, and identifies and manages potential threats.
Thus, in this section, we perform a threat analysis of \ida's QES creation flow. 
First, we describe our threat analysis methodology; followed by our results and extracted requirements.



\subsection{Threat Analysis Methodology}
\label{sec-threat-modelling_sub-methodology}

In the last two decades, various threat-modeling methods have been developed, including several use-case studies~\cite{DBLP:conf/uss/ThompsonMPV24}.
While no single threat-modeling standard exists, the \owasp~\cite{OWASP_threat_modeling_process} suggests several practical models.
We leverage \strideLink~\cite{shostack2014threat,STRIDE_Microsoft} for identifying threats and a \dreadLink-like~\cite{shostack2014threat} model for managing threats.

\tbf{\stride} is a structured security-threat-identification model originally developed by Microsoft.
\stride categorizes threats into six types: \lbrk\emph{\tbf{S}poofing}, \emph{\tbf{T}ampering}, \emph{\tbf{R}epudiation}, \emph{\tbf{I}nformation Disclosure}, \emph{\tbf{D}enial of Service}, and \emph{\tbf{E}levation of Privilege}.
A \stride threat analysis starts from a \acf{dfd}, depicting how relevant data flows through the given system.
Particularly, the \dfd highlights trust boundaries and interactions between them.
For each element, such as data store or data flow, one then checks which threat category is applicable.
\stride's outcome is a structured list/matrix of potential threats, which can then be managed (assessed, prioritized, and addressed). 

\tbf{DREAD} (\emph{\tbf{D}iscoverability, \tbf{R}eproducibility, \tbf{E}xploitability, \tbf{A}ffected users, \lbrk\tbf{D}amage}) is an exemplary methodology to assess threats. 
However, since original DREAD has been essentially abandoned in practice due to its complexity, we use a modified variant inspired by Koch et al.~\cite{DBLP:conf/primelife/KochKPR20} using impact and likelihood to evaluate a threat's priority.

\tbf{Overarching}, we use the four-question framework as suggested by the \lbrk Threat Modeling Manifesto and \owasp:
      \tdFin{protocol-like frame for the "analysis flow" :)}
      \td{arXiv: itemize for, e.g., PRIOritities}
\begin{enumerate}
	\item[\tbf{1)}] \tbf{Scope:} \quoteInline{What are we working on?} \RarrB \ida's QES Creation Flow.

	\item[\tbf{2)}] \tbf{Threat Identification:} \quoteInline{What can go wrong?} \RarrB \strideLink model.
  \begin{enumerate}[label={\alph*)}, font={\bfseries}]
    \item \tbf{Craft:} \dfd of \ida's QES creation process.
    \item \tbf{Map:} \dfd components to STRIDE categories \rarr identifies threats.
  \end{enumerate}

	\item[\tbf{3)}] \tbf{Threat Management:} \quoteInline{What...to do about it?} \RarrB DREAD-like model.
  \begin{enumerate}[label={\alph*)}, font={\bfseries}]
    \item \tbf{Assess Threats:} evaluating with the following scales:
        \\\tbf{Impact:} \tit{(1\rarr{negligible}, 2\rarr{medium-low}, 3\rarr{medium-high}, 4\rarr{severe})},
        \lbrk\tbf{Likelihood:} \tit{(1\rarr{low}, 2\rarr{medium-low}, 3\rarr{medium-high}, 4\rarr{high})}.
    \item \tbf{PRIOritize Threats:} computing priority score\\ \tit{(Impact.nr $\times$ Likelihood.nr)} and map to the corresponding priority group: 
        \cboxRqBacklog{\tbf{High}} \tit{(11\ldots16} \& \emph{Impact=4 or Likelihood=4 regardless of priority score)},
      \cboxRqBacklog{\tbf{Medium}} \tit{(6\ldots10)}, or 
      \cboxRqBacklog{\tbf{Low}} \tit{(1\ldots5)}. 
    \item \tbf{Check MITigation:} assessing current means of mitigation:\\
      \tit{Good-Enough (GE), \cboxMIT{Needs-Improvement (NI)}, Out-of-Scope (OoS)} 
    \item \tbf{Extract Requirement Challenges:} utilizing priority and current \lbrk mitigation to map each threat to the corresponding requirement group:
      \ifnum\IconsInText=1
        \\\cboxRqMust{\tbf{(1)}}\hspace{-0.25em}\ti{\cboxRqBacklog{PRIO.High}\hspace{-0.5em}\tbf{+}\cboxMIT{MIT.NI}\hspace{-0.5em}} \rarr addressing earliest possible.
      \else
        \\\tbf{(1)} \ti{(PRIO.High+MIT.NI)} \rarr address earliest possible (\enquote{must-consider});
      \fi
      \ifnum\IconsInText=1
        \\\tbf{(2)} \ti{\cboxRqBacklog{PRIO.Medium}\hspace{-0.5em}\tbf{+}\cboxMIT{MIT.NI}\hspace{-0.5em}} \rarr good to have in mind.
      \else
        \\\tbf{(2)} \ti{(PRIO.Medium+MIT.NI)} \rarr good to have in mind (\enquote{be-aware});
      \fi
      \ifnum\IconsInText=1
        \\\tbf{(3)}	the rest \ti{(\cboxRqBacklog{PRIO.Low}, MIT.GE, MIT.OoS)} \rarr on the radar.
      \else
        \\\tbf{(3)} \ti{(PRIO.Low, MIT.GE, MIT.OoS)} \rarr on the radar (\enquote{in-backlog}).
      \fi
  \end{enumerate}

	\item[\tbf{4)}] \tbf{Evaluate:} \quoteInline{Did we do a good job?} \RarrB cf. \lnRef{sec-conclusion}.
\end{enumerate}

\paragraphBold{Re-Iteration Note}
Using these structured approaches, we may uncover threats that otherwise remain unseen. 
Following, they help us to to extract requirement challenges for the QES-creation flow. 
Besides our primary requirements, we have also noted less important/urgent requirements; e.g., for later consideration.
\lbrk Generally, it is good practice to regularly re-check and iterate over all threats.







\subsection{Threat Analysis Results and Requirement Challenges} 
\begin{figure}[htb]
 \centering
 \resizebox{\textwidth}{!}{
   \includegraphics{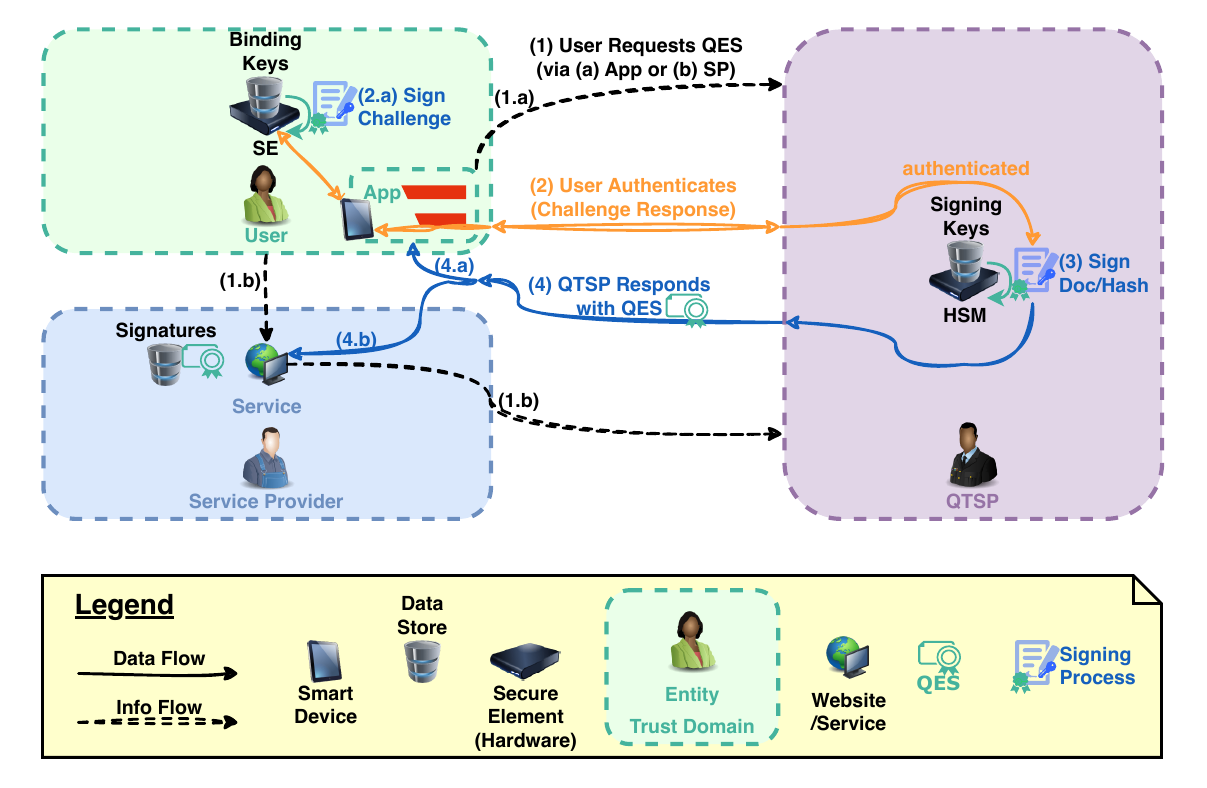}
 }
 \caption{QES-creation Data Flow Diagram within \ida (essential parts). 
 }
 \label{fig::DFD::QES-creation-IDA}
\end{figure}

\parB{Threat identification using the STRIDE model}
\Cref{fig::DFD::QES-creation-IDA} shows our \dfd for \ida's QES-creation flow, focusing on the most essential components.
We omit \tit{standard} steps, such as clicking a button, as well as flows affecting only a single entity.
To better highlight relevant data transitions between entities, we introduced the \emph{Info Flow} as distinguisher from \emph{Data Flows}.
For instance, a QES-creation request is an \emph{Info Flow}, while a QES response including a signature is considered a \emph{Data Flow}.
Further, we assume that a user has already registered at \ida and that all components have established key material.

Now we check for each \dfd component the six STRIDE categories.
To keep the scope of the paper, we focus on QES-relevant threats.
For instance, \tit{standard} threats like integrity of sent/stored data might be briefly mentioned, but are out-of-scope regarding \enquote{must-consider} system requirements.

\parB{Threat management using our DREAD-like model}
Based on the \lbrk initial threat-identification matrix, we assess each threat and compute its priority score.
Utilizing the priority and current mitigation, we map each threat to the corresponding requirement group.
Together with the mentioned focus on scope, we extract primary \faMountainText requirement challenges (RCs) in three areas: \todo{reduce listing if space needed}
\newpage
{
  \ifnum\IconsInText=1
    \renewcommand{\labelitemi}{\faMountainText}
    \renewcommand{\labelenumi}{\faMountainText}
  \else
    \renewcommand{\labelenumi}{\textbf{(\theenumi)}}
  \fi
  \begin{enumerate} 
    \item \cboxRqMust{\ifnum\IconsInText=1\tbf{(1)}\fi \faCogsText{}\textbf{Distributed Service Robustness}}(including \faKeyText{}Split Keys)
    \item \cboxRqMust{\ifnum\IconsInText=1\tbf{(2)}\fi \faRunningText{}\textbf{Agile Crypto Deployment}}(\faExchangeText Crypto Agility + \faAtomText{}PQ Security)
    \item 
    \ifnum\IconsInText=1\cboxRqMust{\ifnum\IconsInText=1\tbf{(3)}\fi \faHandsHelpingText{}\textbf{Active User Involvement}}(\faUserFriendsText{}User Participation + \faBookText{}User-accessible 
      \\\makebox[26.5em][l]{ }Signing Logs).
    \else
      \textbf{Active User Involvement} (User Participation 
      \\\makebox[13.0em][l]{ }+ User-accessible Signing Logs).
    \fi
  \end{enumerate}
}
For enhanced readability, we present these RCs together with potential \lbrk privacy-preserving technologies and practical aspects in~\Cref{sec-qoes-concept::sub-rq-ppc-mapping}. 
Further, we show the full threat analysis matrices of the \dfd components in~\Cref{appx-full-threat-matrix}.


\ifnum\IconsInText=1
	\section{\faYinYang\nobreakspace \qoesign Concept}
\else
	\section{\qoesign Concept}
\fi
\label{sec-COLLiDE-concept}
This section (i) describes relevant privacy-preserving (collaborative-computation) (P2C2) technologies and (ii) maps them to our extracted requirement challenges, along with legal and practical-efficiency aspects.
Then, (iii) we outline two \lbrk specific \qoesign instantiations, offering different tradeoffs.

\subsection{\faShieldText Privacy-Preserving Collaborative Computations (P2C2)}

In the last two decades, primarily three practically relevant privacy-preserving computation technologies emerged, which fit our QES-creation use case~\cite{DBLP:journals/ieeesp/Smart23_COED}:
\lbrk(a) {\thresholdCryptoArXiv} or {\mpcArXiv} in general, (b) {\heArXiv}, and (c) trusted/secure hardware, such as a {\hsmArXiv} or {\teeArXiv}.

In the following, we refer to HE and MPC as \enquote{Virtual Security Module} (VSM), highlighting the secure-computation similarity of an HSM in software.

\parB{\faShieldText MPC}
Is a (software-based) cryptographic building block, which enables a set of parties to jointly compute a function in a private manner.
MPC ensures privacy of secret party inputs and intermediate values.
All parties or a dedicated subset receive the output values.

General-purpose MPC frameworks support virtually any computation with a dynamic amount of participating parties, such as \mpsz~\cite{mp-spdz} or \mpyc~\cite{MPC-Engine:MPyC-Website}\tdFin{reduce refs if space needed}.
Web-based MPC frameworks even enable parties to join MPC computations via the web browser, such as the \wsz~\cite{DBLP:conf/cans/BuchsteinerKRR25}
or the \mpyw~\cite{MPC-Engine:MPyC-Web-GH}.
These recent advances smoothen the entrance to MPC and practically enable end users to participate in MPC computations.


\parB{\faShieldText HE}
Is another (software-based) cryptographic building block, which directly computes on encrypted data.
The user creates a homomorphic ciphertext of the document/hash, which is then \enquote{evaluated} by the \qtsp without getting to know the underlying data.
Then, the user decrypts this \enquote{evaluated} ciphertext to receive the plain signature.
Plain HE offers a different trust model since usually only one external computing party is involved. 
Certain use cases even combine MPC and HE; e.g., encrypting data using HE and splitting the evaluation key using MPC.

\parB{\faShieldText HSM \& software-hardware hybrids}
HSMs provide secure and private \lbrk computations within a hardware component.
While plain HSMs offer usually higher security guarantees due to dedicated hardware, they lack cryptographic agility and only offer computations by one party (non-P2C2 technology).

In terms of combining HSMs with collaborative computation, e.g., the \lbrk company \ttt{i4p} provides \enquote{TRIDENT Multi-Party HSM}~\cite{i4p_trident_mpc}.
This Multi-Party HSM (MP-HSM) approach enables, e.g., distributed RSA encryption and signing. 
\lbrk Each participating party has only a key's share and performs MPC\lbrk computations within their HSM.
Parties can even distributedly create the key shares, so that the full key never resides at any party throughout the key's lifecycle.
Moreover, TRIDENT supports \eidas-compliant QESs.

In terms of general \teeP, e.g., Google provides MPC computations along cloud-hosted \teeP and blockchain~\cite{web_Google_mpc_confidential_space}.\\



\subsection{Mapping of P2C2s to Requirements \& Practical Aspects}
\label{sec-qoes-concept::sub-rq-ppc-mapping}

\Cref{tab-qoes::rq-p2c2-mapping} shows the mapping of our selected P2C2 technologies, and the current \ida solution using an HSM, to our extracted requirement challenges (RCs) and practical aspects (legal requirements and practical efficiency).

\tbf{$\Sigma$ All in all.}
As of now, no P2C2 technology matches all RCs and practical aspects simultaneously.\tdFin{Swiftly switching to another deployed cryptographic protocol with critical keys never being fully at one destination (e.g., \qscd), enhances the system's overall crypto deployment strategy; noticeably improving the system's security \& privacy.}
For instance, while the current solution of \ida (HSM) meets all legal requirements and is practical enough, it does not fulfill our RCs; marking it as potentially insecure for future developments.
At the other end of the spectrum, VSMs provide the highest flexibility, e.g., VSM-MPC meeting all extracted RCs, but \emph{currently} lack practical efficiency and a legal implementation process (ensuring specific (ISO) standardizations etc.). 

In time, VSMs might fulfill all requirements.
In general, software-hardware hybrid solutions seem to be the best choice in the long run.
These hybrids \lbrk potentially offer Distributed Robustness, Crypto Agility, User Participation, and hardware-backed security simultaneously.
Such as the Multi-Party HSM, which has the potential to fulfill all requirements too when overcoming the agility and user participation RC.

In the following, we describe each RC and practical aspect in more detail and evaluate the different P2C2 technologies on them.\\


{


  \newcolumntype{P}{>{\centering\arraybackslash}m{.16\textwidth}}

  \setlength{\heavyrulewidth}{0.08em}     
  \setlength{\belowrulesep}{0ex}        
  \setlength{\aboverulesep}{0ex}        


\begin{table}[htb]
  \caption[Mapping of selected P2C2 to requirements and practical aspects]{ 
    Our selected P2C2 (columns) mapped to extracted requirements and practical aspects (rows).
    The three types of \tbf{mapping circle} constitute: 
    \lbrk\circFullTab{circleGreen} {\color{circleGreen}\tbf{full (green):}} full support/practical enough; 
    \\\circHalfBotTab{enumBlue} {\color{enumBlue}\tbf{half-full (blue):}} support/practicality depends on concrete instantiation;
    \\\circEmptyTab{circleGray} {\color{circleGray}\tbf{empty (gray):}} \emph{currently} not practical enough;
    \\\circCrossRed{circleRed}{-1} {\color{circleRed}\tbf{cross (red):}} no support.
    \\\enquote{*} marks the current solution of \ida. 
  } 
\vspace{1em}
\centering
\resizebox{\textwidth}{!}{
\begin{tabular}{m{.2\textwidth}
    *{5}P
  }

  \multirow{2}{.2\textwidth}{\shortstack[l]{\tbf{Requirement}\\\tbf{Challenges} \faMountainText}} & \mc{2}{c}{\faShieldText \tbf{In Hardware}} & \mc{3}{c}{\faShieldText \tbf{Pure Software (VSM)}} \tn
  \cmidrule(lr){2-3} \cmidrule(lr){4-6}
                            & HSM*         & MP-HSM                     & HE            & MK-HE     & MPC \tn
  \toprule 
  \vspace{0.2em}\shortstack[l]{\faCogsText Distributed\\Robustness} & \cHalf & \cFull & \cHalf & \cYes & \cYes \tn 
  \vspace{0.2em}\shortstack[l]{\faKeyText Split\\Keys} & \cNoRed   & \cYes & \cNoRed &  \cYes & \cYes \tn
  \midrule
  \vspace{0.2em}\shortstack[l]{\faExchangeText{}Crypto\\Agility } & \cNoRed & \cNoRed & \cYes     & \cYes & \cYes \tn
  \vspace{0.2em}\shortstack[l]{\faAtomText PQ\\Security} & \cHalf & \cHalf & \cYes & \cYes & \cYes \tn
  \midrule 
  \vspace{0.2em}\shortstack[l]{\faUserFriendsText User\\Participation} & \cNoRed & \cHalf & \cYes   & \cYes & \cYes\tn
  \vspace{0.2em}\shortstack[l]{\faBookText User-access.\\Signing Logs} & \cYes & \cYes & \cYes   & \cYes & \cYes\tn
  \bottomrule \toprule 
  \vspace{0.2em}\shortstack[l]{\tbf{Legal Requirements} \faBalanceScaleText} & & & & & \tn
  \vspace{0.2em}\shortstack[l]{\faFileSignatureText{}AdES\\(sole control ...)} & \cYes & \cYes & \cYes  &  \cYes & \cYes \tn
  \vspace{0.2em}\shortstack[l]{\faAwardText QES\\(+ QSCD ...)} & \cYes & \cYes & \cHalf {\color{enumBlue}\tbf{?}} &  \cHalf {\color{enumBlue}\tbf{?}} & \cHalf {\color{enumBlue}\tbf{?}} \tn
  \bottomrule \toprule 
  \vspace{0.2em}\shortstack[l]{\tbf{Practical}\\\tbf{Efficiency} \faStopwatchText} & \cYes & \cYes & \cNo & \cNo & \cHalf \tn
  \bottomrule
  \end{tabular}
} 
  \label{tab-qoes::rq-p2c2-mapping}
\end{table}
}


\ifnum\IconsInText=1
	\noindent\cboxRqMust{\tbf{\faMountainText{}RC-1 \faCogsText{}Distributed Service Robustness}}\\
	\noindent\cboxRqMust{\tbf{\faMountainText{}RC-1-a \faCogsText{}Distributed Robustness.}}
\else
	\noindent\underline{\tbf{(RC-1) Distributed Service Robustness}}\\
	\noindent\tbf{(RC-1a) Distributed Robustness.}
\fi
\ttt{Definition:} To enhance system \lbrk robustness (availability \& stability), the system distributes its QES creation \lbrk so that a \qscd/entity can drop \emph{during} the computation process without causing service delays or outages.

Technically, the \qtsp's server represents a single point of failure; e.g., in case of denial-of-service-attacks.
Especially for government-related services that rely on QESs, availability is of paramount importance to ensure continuous operation. \\

\cboxPC{\ttt{\faShieldText P2C2s:}}
In general, backup servers/\qtspP can provide fallback signing; \lbrk enabling a weaker variant of system robustness.
However, if, e.g., a whole \qtsp is not available, users might need to manually switch to another \qtsp/service provider, causing service delays.
Moreover, this \enquote{static fallback approach} \lbrk violates the Split Keys RC since a \qtsp/\hsm needs to have the full private signing key for providing the service (see (RC-1b) below). \\
\indent To achieve our definition of distributed service robustness, one usually \lbrk leverages threshold cryptography via, e.g., threshold secret sharing.
Threshold computations require $\geq t$ \emph{out of} $n$ \qtspP/instances, the $threshold$, and are supported by MP-HSM, MK-HE, and MPC.
This threshold value can be set \lbrk individually; e.g., from $\mathit{5}$ \qtspP, we might require $\mathit{\geq 3}$.
While the setup is \lbrk heavier as various \qtspP are involved, then for the QES creation, \emph{some} \qtspP can drop without affecting the service's functionality.\\


% ...

\newpage
\ifnum\IconsInText=1
	\noindent\cboxRqMust{\tbf{\faMountain{}RC-1-b \faKey{}Split Keys.}}
\else
	\noindent\tbf{(RC-1b) Split Keys.}
\fi
\ttt{Definition:}
To enhance system security, the private signing key never resides fully at one \qscd/entity.

Given the signing key, everyone can craft a signature \enquote{on behalf of} the affected user.
Hence, due to the possibility of data leakage, and utilizing the Distributed Robustness RC, a \qscd/entity shall only have a key share.

\cboxPC{\ttt{\faShieldText P2C2s:}}
The single-instance computing approaches HSM and HE need the full key.
The threshold-computation approaches provide this \enquote{key split} by design.\\

\ifnum\IconsInText=1
	\noindent\cboxRqMust{\tbf{\faMountain{}RC-2 \faRunning{}Agile Crypto Deployment.}} 
	\\\noindent\cboxRqMust{\tbf{\faMountain{}RC-2-a \faExchange*{}Crypto Agility.}}
\else
	\noindent\underline{\tbf{(RC-2) Agile Crypto Deployment}} 
	\\\noindent\tbf{(RC-2a) Crypto Agility.}
\fi
\ttt{Definition:}
To ensure system security in an ever-changing world, the system provides Crypto Agility (transition \emph{smoothly and swiftly} from one cryptographic protocol to another).

In July 2025, NIST published a whitepaper regarding the importance of Crypto Agility along the constant needs of transition: \enquote{Given the long-desired lifespan of many devices, it is generally more cost-effective to design the \lbrk device for such transitions during its development}~\cite{NIST-crypto-agility_2025_Barker_et_al}.
For instance, transitioning from an RSA-based to a lattice-based (PQ-secure) signature protocol without impediments and noticeable service shutdowns in a relatively short time \lbrk period.
Additionally, given a general breakthrough cryptanalytical result, a static \lbrk solution endangers the deployed system.

\cboxPC{\ttt{\faShieldText P2C2s:}}
Crypto Agility is given for all VSMs since a software module can be replaced relatively easily.
A hardware-based solution would need to offer an option to change the used protocol via, e.g., firmware updates, which usually is not possible for HSMs.
Thus, HSMs would require manufacturing new hardware, which is more costly in terms of time.

Further, a hybrid solution could leverage VSM computations within hardware.
Then, even if respective algorithms are compromised, VSM provide agility.\\

\ifnum\IconsInText=1
	\noindent\cboxRqMust{\tbf{\faMountain{}RC-2-b \faAtom{}PQ Security.}}
\else
	\noindent\tbf{(RC-2b) PQ Security.}
\fi
\ttt{Definition:}
To ensure system security in the quantum computation era, the system provides PQ security.

In the advent of practical quantum computers, most asymmetric encryption and signing schemes, respectively, can be practically broken.
For instance, \lbrk RSA-based or discrete-logarithm-based schemes using Shor's algorithm~\cite{DBLP:conf/focs/Shor94}.

\cboxPC{\ttt{\faShieldText P2C2s:}}
Some modern \hsmP or MP-HSMs, such as \enquote{Luna HSM}~\cite{Luna_HSM} or TRIDENT MPC~\cite{i4p_trident_mpc} respectively, are an intermediate case since they provide a PQ-secure signing algorithm, but lack agility, endangering the system if this specific algorithm gets compromised.
Both MPC and (MK-)HE can be instantiated using PQ-secure algorithms.
In general, if the system provides Crypto Agility, (sustainable) PQ Security is fairly easily achievable by switching to another protocol.\\


\ifnum\IconsInText=1
	\noindent\tbf{\cboxRqMust{\faMountain{}RC-3 \faHandsHelping{}Active User Involvement.}}\\
	\cboxRqMust{\tbf{\faMountain{}RC-3-a \faUserFriends{}User Participation.}} 
\else
	\noindent\underline{\tbf{(RC-3) Active User Involvement}}\\
	\tbf{(RC-3a) User Participation.} 
\fi
\ttt{Definition:} To enhance system trust \& transparency, a user actively participates in the QES-creation computation.

In case of a compromised server, the signing service/QTSP might create QESs without user permission since the key resides \tit{fully} within the server's \hsm \lbrk(for the current solution).
Even worse, affected users might not notice the unauthorized usage. 
When a user participates in the signing process, which should happen anyway only if a user requests it, the risk of unauthorized signatures is reduced.
Moreover, this user participation strengthens the \enquote{sole control} \lbrk requirement on a technical level (cf. legal requirements below).

\cboxPC{\ttt{\faShieldText P2C2s:}} A single-HSM solution solely operates at the \qtsp's server side.
For MP-HSM, we would need to establish a hybrid solution; where \qtspP compute within an HSM, while a user participates with their own device.
However, we are not aware of such a hybrid solution at this point in time.
For HE, while the user is not part of the signing process itself, the user is part of the overall QES-creation process as the \qtsp needs the user-generated HE ciphertext. 
For MPC, the user is actively part of the signing process itself.
To ensure active user participation, we can design the MPC protocol so that the user must be part of the computation, while only up to $threshold-t$ \qtspP' are required.

However, this RC requires corresponding key management on the user's side.\\


\ifnum\IconsInText=1
	\noindent\cboxRqMust{\tbf{\faMountain{}RC-3-b \faBook{}User-accessible Signing Logs.}}
\else
	\noindent\tbf{(RC-3b) User-accessible Signing Logs.}
\fi
\ttt{Definition:} To further enhance system trust \& transparency, a user can verify a server's/\qscd's Signing Logs.

\cboxPC{\ttt{\faShieldText P2C2s:}} Signing Logs can be established for each solution; e.g., establishing a QES-creation blockchain for each user.
Then, each user can check \lbrk via the respective blockchain which signatures have been created.
However, in case of a compromised server such logs might be suppressed.
Thus, combining User-accessible Signing Logs with User Participation provides the highest \lbrk(Active) User Involvement, leading to increased system trust \& transparency.

\subsection{\faBalanceScaleText Legal QES Requirements and \qoesign's P2C2 Technologies}\tdFin{arXiv: subsections}

\noindent For crafting {\qessLinkArXiv}, a device/entity first must meet requirements for an {\adesLinkArXiv}, followed by additional requirements; both outlined in the \eidas regulation~\cite{eIDAS} (cf. Art 3 para 11 and 12).
Further, the EU's NIS-2 directive requirements must be observed; its national transposition/implementation in, e.g., Austria being pending.



\parB{\faFileSignatureText AdES requirements}
An AdES must be (i) uniquely linked to the signatory, (ii) capable of identifying the signatory, and (iii) linked to the signed data ensuring integrity. 
Further, the signatory must be able to use the electronic signature creation data with high level of confidence under their \enquote{sole control}, specified in the CEN EN norm 419 241-1.
While \enquote{sole control} can be achieved on a contractual basis, as with the current HSM solution, with the P2C2 technologies we can achieve this property even on a computational level when the user participates; strengthening the fulfillment of this requirement.
From a general technical viewpoint, all P2C2 technologies should be able to fulfill the AdES requirements.

\parB{\faAwardText Additional QES requirements}
A QES must be created (i) by a \qscd based on (ii) a qualified certificate.
Generally, \eidas defines \qscd as a \quoteInline{configured software or hardware used to create an electronic signature}. 
To be classified as a \qscd, a device must meet \eidas' Annex II requirements.
For eSignature certificates, \eidas generally defines them as an \quoteInline{electronic attestation which links electronic signature validation data to a natural person and confirms at least the name or the pseudonym of that person}. 
Qualified certificates must (a) meet \eidas' Annex I requirements and (b) issued by a \qtsp, which is defined as \quoteInline{a trust service provider who provides one or more qualified trust services and is granted the qualified status by the supervisory body}.

The qualified certificate should be creatable for all P2C2 technologies; e.g., by an existing \qtsp.
Achieving a \qscd status for user devices is the most challenging part.
The \qscd part needs deeper legal and technical investigations; e.g., checking required (ISO) certificates for the legal implementation process.

\parB{NIS-2 directive}
Defines general cybersecurity requirements; especially for \lbrk essential entities, such as \qtspP (cf Art 3 para 1)\fn{Directive (EU) 2022/2555 on measures for a high common level of cybersecurity across the EU, amending Regulation (EU) 910/2014 and Directive (EU) 2018/1972 and repealing Directive (EU) 2016/1148 (NIS-2 Directive)}.
In October 2024, the EU adopted an implementing regulation for this directive\fn{Implementing Regulation (EU) 2024/2690, specifying application rules for \lbrk Directive (EU) 2022/2555}. 
For instance, the \lbrk regulation further specifies NIS-2's cryptography requirements from Article 21:
\begin{quote}
	\quoteInline{the protocols or families of protocols to be adopted, as well as cryptographic algorithms, cipher strength, cryptographic solutions and usage practices to be approved and required for use in the relevant entities, following, where appropriate, a \tbf{cryptographic agility approach}.}
\end{quote}
However, e.g., the \emph{currently} deployed \ida solution, single (static) HSM, does not provide this agility.
Thus, leveraging \qoesign's crypto-agile P2C2 technologies (VSMs/adapted hybrids) strengthens the fulfillment of this EU \lbrk implementing regulation for the QES-creation process.
\subsection{\faStopwatchText Practical Efficiency of \qoesign's P2C2 Technologies}
Since \ida (currently) bases its QES creation on HSMs (non-P2C2), \lbrk we assume the efficiency is practical enough.
For MP-HSMs, we assume its \lbrk practical enough too as it builds upon HSMs; especially given that \lbrk\enquote{TRIDENT MPC} already provides this solution.

Fabianek et al.~\cite{DBLP:journals/corr/abs-2410-16442} investigated MPC and HE in general \enquote{data space} use cases.
They concluded that both MPC and HE are promising technologies, yet need further research and practical improvements to be fully integrated. 
On a meta level, also \eid services such as QESs fall into these investigations. 
More specifically, HE for signing within the encrypted document/ hash is quite impractical at the moment.
MPC provides security levels primarily based on (i) adversary corruption (passive vs. active) and (ii) nr. of adversaries \lbrk(honest vs. dishonest majority).
The lower the chosen security level, the faster the protocol run; and vice-versa.
For practical usage, we can leverage MPC's Crypto Agility and tune the security level based on the system's needs.
For instance, we can choose the max. security and accept the slower runtime for documents/data that require a QES.
While for general cases, we might default to security against active adversaries for an honest majority; striving for an AdES.
Further, MPC's performance usually grows quadratically with the \lbrk nr. of participating parties.
Thus, we recommend a range of three to five \qtspP for \emph{initial experiments}, with accompanying test benchmarks.

From our investigations and use case, HE seems less performant compared to MPC.
Hence, MPC and its hybrid versions are \emph{currently} the most promising P2C2 technologies for QoeSiGN's QES-creation process.

\begin{figure}[htb]
  \centering
  \resizebox{\textwidth}{!}{
    \begin{subfigure}{0.5\textwidth}
      \includegraphics[width=\linewidth]{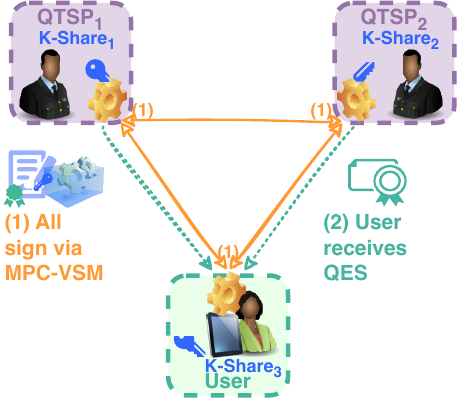}
      \caption{MPC-VSM}
      \label{fig-QoeSiGN-concept_MPC-VSM}
    \end{subfigure} 
    \hspace{.5em}
    \vline
    \hspace{.5em}
    \begin{subfigure}{0.5\textwidth}
      \includegraphics[width=\linewidth]{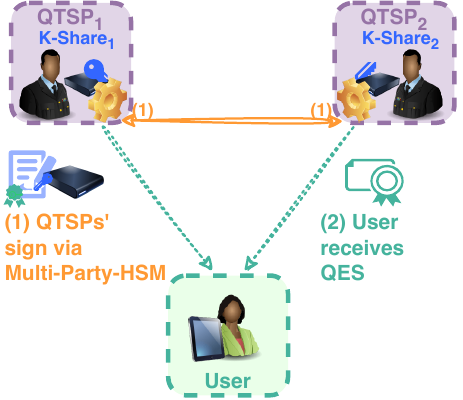}
      \caption{Multi-Party HSM}
      \label{fig-QoeSiGN-concept_MP-HSM}
    \end{subfigure}
  }
  \caption{Two \qoesign variants, offering different properties and security tradeoffs.}
  \label{fig-QoeSiGN-instantiation}
\end{figure}

\ifnum\IconsInText=1
	\subsection{\faYinYang\nobreakspace Different \qoesign Instantiations for Respective Needs}
\else
	\subsection{Different \qoesign Instantiations for Respective Needs}
\fi
Since the \qoesign concept explored several potentials, we can instantiate it using, e.g., one of the investigated P2C2s depending on the respective needs.
For the QES-creation process, and \eid systems in general, we balance a tradeoff between maximum security, (distributed) robustness, (crypto) agility, and user involvement.
\Cref{fig-QoeSiGN-instantiation} shows two such concrete instantiations.
To simplify, we assume a user directly receives the QES from the \qtsp (instead of via an SP).

\parB{MP-HSM \qoesign for distributed secure hardware}
In this variant, \lbrk several \qtspP jointly compute the QES within an HSM, utilizing MPC for the distribution.
MP-HSMs provide hardware-based security with enhanced \lbrk (distributed) robustness.
Setting up the users' key shares and establishing \lbrk secure communication channels for the joint computation constitute the \lbrk primary challenge in this \qoesign variant.
However, this variant could already be used since there exist \eidas-compliant MP-HSM solutions.

\parB{VSM-MPC \qoesign for highest agility and user involvement}
In this variant, several \qtspP and the user jointly compute the QES via MPC; \lbrk providing the highest crypto agility and user involvement.
Regarding the private signing key shares, a user needs to be part of the key creation process when registering at the system.
Certifications of user devices, for being an eligible participation/\qscd device, is the hardest challenge in this variant.
We \emph{might} need to check which phone models are eligible to actively participate.
Hence, in this \qoesign variant, we must ensure an applicable user-device model and \lbrk status.
For non-\qscd-compliant devices, we could use, e.g., a weaker VSM-MPC security protocol; providing an AdES, which \emph{can} still provide legal value.



\newpage


\section[Conclusions]{Conclusions and Further Work}
\label{sec-conclusion}


\noindent
In this paper, we addressed the single-point-of-failure aspect within the (remote) QES-creation process.
Although we looked at the specific case of \ida, our identified threats and extracted requirement challenges can be transferred to the general QES ecosystem. 
As such, we developed the \ifnum\IconsInText=1\faYinYangText\nbrSp\fi\qoesign concept by utilizing novel privacy-preserving collaborative computation (P2C2) technologies. 
While \emph{currently} no P2C2 technology addresses all requirements and practical efficiency simultaneously, we presented different instantiation possibilities for respective needs.
For instance, the MP-HSM software-hardware hybrid for distributed HSM computations among \qtspP; or the MPC-VSM for max. crypto agility und user involvement, where the user even participates in the QES computation.

\paragraphBold{Impact on the current QES ecosystem}
From a deployment point of view, \qtspP would first need to adapt \qscd; e.g., supporting Multi-Party \hsmP or instantiating a pure software-based VSM.
These \qtsp-related adaptations might require additional audits to ensure the needed ISO certifications.
\lbrk Additionally, collaborative-computation \qtspP' must setup trusted and \lbrk reliable communication channels between them.
\sepP might \emph{only} need to change the signature request to invoke all involved \qtspP.

The biggest challenge constitutes the active computation participation of user devices, leaving room for further research and implementation investigations; primarily regarding certifications and practical security and efficiency \lbrk aspects.
Especially since \qoesign is a (theoretical) concept, performing real-life implementation benchmarks would be required to bring it further into practice.

\newpage
\paragraphBold{Full eID system view}
In addition to the QES-creation process, we can \lbrk consider the authentication and registration process; including, e.g., setting up cryptographic keys.
Since the QES creation relies on properly established key material, registration is a critical part of an QES/eID system. 
This key setup needs specific focus for the user-participation scenarios.


\parB{\quoteInline{Did we do a good job?}, legal upgrades \& threat modeling by design}
To answer the evaluation part of \owasp's four-question model, we would need to further implement the suggested system enhancements in the QES ecosystem, including all stakeholders.
Though, as a concrete next step, we can build upon {\kerckhoffLinkArXiv} and collectively work on \qoesign's concept; e.g., iterating over the threat analysis with other experts from the different domains.
Further, in addition to a security-related threat analysis, we can consider a \lbrk privacy analysis too, such as the LINDDUN~\cite{LINDDUN-intro_deng2011privacy,LINDDUN-cont_sion2025robust} methodology. 
Moreover, the legal framework around QES, and critical infrastructure in general, could further investigate our identified primary \enquote{meta requirements} (distributed robustness, user involvement, and agile (crypto) deployments).

For instance, in the medical domain, \quoteInline{regulators and industry standards recognize the need for 'secure by design approaches' [..] and now require threat modeling for medical devices [..]}~\cite{DBLP:conf/uss/ThompsonMPV24} as these may be critical for people's safety \& wellbeing.
Since eSignatures are an essential component in today's digitized world, requiring a rigorous threat modeling for this domain would be one of the next steps to enhance the security, privacy, and reliability of the (QES) \lbrk eSignature ecosystem.
Especially in our ever-changing world, protecting the \lbrk security \& privacy of individuals is of utmost importance, adhering to the human rights and contributing to safety \& wellbeing. 

\tdi{ACK + Disclosure of Interests from LNCS Guideline}


\hspace{0pt}\\
\begin{credits}
\subsubsection{\ackname}
This work received funding from 
\href{https://www.kiras.at/en/financed-proposals/detail/prepared-post-quanten-sichere-eid-fuer-oesterreich/}{\textbf{PREPARED}}, a project funded by the Austrian security research programme KIRAS of the Federal Ministry of Finance (BMF)
and \href{https://cordis.europa.eu/project/id/101168311}{\textbf{LICORICE}}, a European Union's Horizon Europe project (no. 101168311). 
And we thank the anonymous reviewers for their valuable feedback.



\tbf{Credits.}
\Cref{fig::DFD::QES-creation-IDA} and~\Cref{fig-QoeSiGN-instantiation} drawn using \ttt{draw.io}; key symbol from \lbrk\ttt{IconFinder.com}.
Table icons from \ttt{fontawesome}.

\end{credits}


\bibliographystyle{splncs04}
\bibliography{_Support/bib_MPC, _Support/bib_b4M, _Support/bib_eID, _Support/bib_PQC, _Support/bib_threat-modeling,  
  _Support/cryptobib/abbrev0, _Support/cryptobib/abbrev1, _Support/cryptobib/abbrev2, _Support/cryptobib/abbrev3
  }\tdFin{update links' access date and bring to nice form :)}

  \newpage
  \appendix
  \setcounter{section}{0}
  

\section{On the Equivalence Principle of \\ QESs and Handwritten Signatures}
\label{appx-QES-handwritten-equivalence}
As mentioned in~\lnRef{sec-Intro}, \eidas establishes the principle that QESs \emph{may be} equivalent to handwritten signatures (cf. Art 25 para 21), depending on the governing law and particularities of the individual case.
As such, the respective governing law \emph{might} specify that, e.g., an advanced electronic \lbrk signature (AdES) or even a simple electronic signature (SES) may substitute for a handwritten signature. 
Likewise, the parties of a contract may agree on such a substitution (provided that no mandatory statutory requirements preclude it).
In contrast to a handwritten substitution/equivalence, the respective governing law, or a contract's parties, \emph{might} specify that no eSignature suffices and requires a handwritten signature. 

Thus, the specific circumstances of each individual case determine the legal effect of the respective signature type. 
For instance, the Austrian Notarial Code accepts any eSignature for official notarial acts that is affixed in the presence of a notary (cf. Art 1a NO).
If neither the respective governing law nor any contractual arrangements prescribe a specific form, the general principle under the eIDAS regulation (Art 25 para 2) and, e.g., Austria's specification (Art 4 SVG) applies, according to which a QES is equivalent to a handwritten signature.

\section{Full Threat Analysis Matrix}
\label{appx-full-threat-matrix}
\Cref{tab-full-threat-matrix-DF} \tit{(Data Flow)}, \Cref{tab-full-threat-matrix-IF} \tit{(Info Flow)}, \Cref{tab-full-threat-matrix-SignP} \tit{(Signing Process)}, and \Cref{tab-full-threat-matrix-DS} \tit{(Data Store)} represent our full threat-analysis matrix, which is computed as described in~\Cref{sec-threat-modelling_sub-methodology}:
\begin{enumerate}
  \item Mapping each \acf{dfd} component to the six \stride \lbrk categories \emph{(\tbf{S}poofing}, \emph{\tbf{T}ampering}, \emph{\tbf{R}epudiation}, \emph{\tbf{I}nformation Disclosure}, \lbrk\emph{\tbf{D}enial of Service}, \emph{\tbf{E}levation of Privilege)}
  \item Then, for each threat:
  \begin{enumerate}
    \item Assessing impact \emph{(I.1-4)} and likelihood \emph{(L.1-4)} 
      \\ to compute the priority score \emph{($I \times L$)}
      \\ and map to a priority group \emph{(High, Medium, Low)} 
    \item Assessing current mitigation 
      \\ \emph{(Good-Enough (GE), Needs-Improvement (NI), Out-of-Scope (OoS))}
    \item Mapping to a requirement group 
    \begin{dingautolist}{182}
      \item \tit{\cboxRqMust{(\enquote{Must Consider})}}
      \item \tit{(\enquote{Be Aware})}
      \item \tit{(\enquote{Backlog})}
      \end{dingautolist}
  \end{enumerate}
\end{enumerate}
Please note that in the following tables \enquote{N/A} denotes that a threat is out of scope for this paper (not applicable) and \enquote{SE} denotes a Secure Element hardware component. 




\tdFin{used acronyms glossary?}
\td{some idea for symbols :D \faShield* Mitigation ..\faChartBar[regular] Priority Range}
\td{\faDumbbell Lifting :D .. \faMap[regular] Requirements Map -> the way to go :D  }


{ 
\newcolumntype{H}{>{\raggedleft\arraybackslash}p{2.5em}} 

\newcolumntype{D}{>{\raggedright\arraybackslash}p{13em}} 

\newcolumntype{A}{>{\centering\arraybackslash}p{5em}} 
\newcolumntype{M}{>{\centering\arraybackslash}p{12em}} 
\newcolumntype{R}{>{\centering\arraybackslash}p{3.5em}} 

\newcommand{\arrDown}{\faLongArrowAltDown}
\newcommand{\arrRight}{\faArrowAltCircleRight}
\newcommand{\cSearch}{\faIcon{search}}

\newcommand{\tabn}{\tabularnewline}

\newcommand{\dfdComp}[1]{\multicolumn{5}{l}{\cellcolor{enumBlue}\color{white}\scriptsize\tbf{#1}}}

\newcommand{\tS}{\scriptsize\emph{S}\cSearch}
\newcommand{\tT}{\scriptsize\emph{T}\cSearch}
\newcommand{\tR}{\scriptsize\emph{R}\cSearch}
\newcommand{\tI}{\scriptsize\emph{I}\cSearch}
\newcommand{\tD}{\scriptsize\emph{D}\cSearch}
\newcommand{\tE}{\scriptsize\emph{E}\cSearch}

\newcommand{\thrDe}[2]{\scriptsize\emph{#1} {#2}}

\newcommand{\thrDeDup}[2]{\multicolumn{4}{l}{\cellcolor{gray!20}\shortstack[l]{\strut\scriptsize\emph{#1}#2\strut}}}

\newcommand{\tdupS}{\emph{\tS}\xspace}
\newcommand{\tdupT}{\emph{\tT}\xspace}
\newcommand{\tdupR}{\emph{\tR}\xspace}
\newcommand{\tdupI}{\emph{\tI}\xspace}
\newcommand{\tdupD}{\emph{\tD}\xspace}
\newcommand{\tdupE}{\emph{\tE}\xspace}

\newcommand{\cPrHigh}[0]{\cellYellow \faExclamationCircle\xspace}
\newcommand{\cPrMed}[0]{\cellBlue \faAdjust\xspace}
\newcommand{\cPrLow}[0]{\cellGray \circEmpty{circleGray}\xspace}

\newcommand{\prH}[3]{\vSpC\cPrHigh \emph{P.High (#3)} \lbr I.#1 \emph{\&} L.#2}
\newcommand{\prM}[3]{\vSpC\cPrMed  \emph{P.Med (#3)} \lbr I.#1 \emph{\&} L.#2}
\newcommand{\prL}[3]{\vSpC\cPrLow  \emph{P.Low (#3)} \lbr I.#1 \emph{\&} L.#2}

\newcommand{\vSpC}{\vspace{-0.9em}\scriptsize}

\newcommand{\assessNA}[1]{\multicolumn{3}{l}{\cellcolor{gray!20}#1}}

\newcommand{\cMiNi}[0]{\cellYellow \faTimesCircle\xspace}
\newcommand{\cMiGe}[0]{\cellGreen \faCheckCircle[regular]\xspace}
\newcommand{\cMiOos}[0]{\cellGray \circEmpty{circleGray}\xspace}
\newcommand{\miGE}[1]{\vSpC\cMiGe  \emph{MIT.GE} \lbr #1}
\newcommand{\miNI}[1]{\vSpC\cMiNi  \emph{MIT.NI} \lbr #1}
\newcommand{\miOS}[1]{\vSpC\cMiOos \emph{MIT.OoS} \lbr #1}

\newcommand{\cRqMu}[0]{\cellOrange\faExclamationTriangle}
\newcommand{\rqMu}{\vSpC\cRqMu \vspace{0.2em} \lbr \scalebox{1.1}{\normalsize\ding{182}}} 

\newcommand{\cRqAw}[0]{\cellBlue\circSmallDot{enumBlue}}
\newcommand{\rqAw}{\vSpC\cRqAw \vspace{0.2em} \lbr \scalebox{1.1}{\normalsize\color{enumBlue}\ding{183}}} 

\newcommand{\cRqBa}[0]{\cellGray\circEmpty{circleGray}}
\newcommand{\rqBa}{\vSpC\cRqBa \vspace{0.2em} \lbr \scalebox{1.1}{\normalsize\color{gray}\ding{184}}}

\begin{table}[thb]
  \caption{Full threat-analysis matrix for the Data Flow (DF) \dfd components.}
	\label{tab-full-threat-matrix-DF}
	\centering
  \footnotesize
  \begin{tabularx}{\textwidth}{HDAMR}
    %
    \scriptsize \emph{ST}\cSearch\linebreak\emph{RI}\cSearch\linebreak\emph{DE}\cSearch 
        & \scriptsize \mr{3}{*}{\emph{Threat Description}} 
        & \scriptsize \faArrowAltCircleRight\linebreak\emph{Priority}\linebreak Imp.\emph{\&}Lh.
        & \scriptsize \faShield*\linebreak\emph{Current}\linebreak\emph{MITigation} 
        & \scriptsize \faMountain\linebreak\emph{Requ.}\linebreak\emph{Group}
        \tabularnewline
    \toprule \addlinespace[-0.0em] 
    \dfdComp{(DF-2) User Authenticates towards QTSP} \tabularnewline
    \addlinespace[-0.19em] \midrule \addlinespace[-0.0em]
    \tS & \thrDe{Authenticate as someone else:}{creating a signature on behalf of someone else}
        & \prH{4}{1}{4!} & \miGE{Cryptographic Binding \& Challenge-response authentication (CRA)} & \rqAw \tabn
    \hhline{~----}
    \tT & \thrDe{Request is modified:}{(a) prohibit a user from logging in; (b) authenticate as someone else}
        & \prL{2}{1}{2} & \miGE{Usually, trusted/known networks and TLS + CRA } & \rqBa \tabn
    \hhline{~----}
    \tR & \thrDe{User signs without being aware of:}{Adversary authenticates as user or triggers authentication via user app, without user noticing it}
        & \prH{4}{1}{4!} & \miGE{Cryptographic Binding + CRA} & \rqAw \tabn
    \hhline{~----}
    \tI & \thrDe{Leak of private binding key}{}
        & \prH{4}{2}{8!} & \miNI{Asymmetric Crypto in Secure Element \Rarr Potential PQ threat} & \rqMu \tabn
    \hhline{~----}
    \tD & \thrDe{User floods QTSP with requests}{}
        & \prL{3}{1}{3} & \miOS{N/A} & \rqBa \tabn
    \hhline{~----}
    \tE & \thrDeDup{Same as (DF-2) \rarr \tdupS}{} \tabn
    \toprule \addlinespace[-0.0em] 
    \dfdComp{(DF-4.a) QTSP responds with QES \rarr User App} \tabn
    \addlinespace[-0.19em] \midrule \addlinespace[-0.0em]
    \tS & \vSpC\emph{Adversary pretends to be QTSP}
        & \prL{2}{2}{4} & \miOS{N/A} & \rqBa \tabn
    \hhline{~----}
    \tT & \thrDe{Request is modified:}{Results in incorrect hash/signature}{}
        & \prL{2}{1}{2} & \miGE{Usually, trusted/known networks and TLS} & \rqBa \tabn
    \hhline{~----}
    \tR & \vSpC\emph{QTSP doesn't respond with QES but saying so}
        & \prL{3}{1}{3} & \miOS{N/A} & \rqBa \tabn
    \hhline{~----}
    \tI & \thrDe{QES leaks}{(if not hash used, full document too? which could leak sensitive information)}
        & \prM{3}{2}{6} & \miNI{Usually, trusted/known networks and TLS \lbr\Rarr Potential PQ threat} & \rqAw \tabn
    \hhline{~----}
    \tD & \thrDe{QTSP floods User with requests}{}
        & \prL{2}{1}{2} & \miOS{N/A} & \rqBa \tabn
    \hhline{~----}
    \tE & \thrDeDup{N/A}{} \tabn
    \toprule \addlinespace[-0.0em] 
    \dfdComp{(DF-4.b) QTSP responds with QES \rarr Service Provider (SP)} \tabn
    \addlinespace[-0.19em] \midrule \addlinespace[-0.0em]
    \tS & \vSpC\emph{Adversary pretends to be QTSP}
        & \prL{2}{2}{4} & \miOS{N/A} & \rqBa \tabn
    \hhline{~----}
    \tT & \thrDeDup{Same as (DF-4.a) \rarr \tdupT}{} \tabn
    \hhline{~----}
    \tR & \vSpC\emph{QTSP doesn't respond with QES but saying so}
        & \prL{3}{1}{3} & \miOS{N/A} & \rqBa \tabn
    \hhline{~----}
    \tI & \thrDeDup{Same as (DF-4.a) \rarr \tdupI}{} \tabn
    \hhline{~----}
    \tD & \thrDe{QTSP floods SP with requests; SP floods User with requests}{}
        & \prL{2}{1}{2} & \miOS{N/A} & \rqBa \tabn
    \hhline{~----}
    \tE & \thrDeDup{N/A}{} \tabn
    \addlinespace[-0.19em] \bottomrule  
  \end{tabularx}
\end{table}

\begin{table}[thb]
  \caption{Full threat-analysis matrix for the Info Flow (IF) \dfd components.}
	\label{tab-full-threat-matrix-IF}
	\centering
  \footnotesize
  \begin{tabularx}{\textwidth}{HDAMR}
    \scriptsize \emph{ST}\cSearch\linebreak\emph{RI}\cSearch\linebreak\emph{DE}\cSearch 
        & \scriptsize \mr{3}{*}{\emph{Threat Description}} 
        & \scriptsize \faArrowAltCircleRight\linebreak\emph{Priority}\linebreak Imp.\emph{\&}Lh.
        & \scriptsize \faShield*\linebreak\emph{Current}\linebreak\emph{MITigation} 
        & \scriptsize \faMountain\linebreak\emph{Requ.}\linebreak\emph{Group}
        \tabularnewline
    \toprule \addlinespace[-0.0em] 
    \dfdComp{(IF-1.a) User requests QES via App \rarr QTSP} \tabn
    \addlinespace[-0.19em] \midrule \addlinespace[-0.0em]
    \tS & \thrDe{Requesting a signature on behalf of someone else}{}
        & \prM{3}{2}{6} & \miGE{Authentication Process: (i) adversary would need to know, e.g., user's phone nr. \& password
        and (ii) affected user gets Auth.request (Challenge-Response) for something the user did not initialize (user should not accept;  thus, small risk of user accepting anyways)} & \rqBa \tabn
    \hhline{~----}
    \tT & \thrDe{Request is modified:}{(a) request fails or (b) reqested for someone else}
        & \prM{3}{2}{6} & \miGE{Usually, trusted/known networks and TLS + following Authentication Process} & \rqBa \tabn
    \hhline{~----}
    \tR & \thrDe{User denies having sent request}{}
        & \prL{2}{2}{4} & \miGE{User gets Authentication Process request} & \rqBa \tabn
    \hhline{~----}
    \tI & \thrDe{Observing that User requests a signature directly at QTSP}{}
        & \prL{2}{2}{4} & \miOS{Usually, trusted/known networks and TLS} & \rqBa \tabn
    \hhline{~----}
    \tD & \thrDe{User floods QTSP with requests}{}
        & \prL{2}{2}{4} & \miOS{N/A} & \rqBa \tabn
    \hhline{~----}
    \tE & \thrDeDup{Same as (IF-1.a) \rarr \tdupS}{} \tabn
    \toprule \addlinespace[-0.0em] 
    \dfdComp{(IF-1.b) User requests QES via Service Provider (SP) \rarr QTSP} \tabn
    \addlinespace[-0.19em] \midrule \addlinespace[-0.0em]
    \vspace{-1.65em} \tS & \thrDeDup{Essentially, same as (IF-1.a) \rarr \tdupS}{\\[-0.45em]\scriptsize(SP requesting for user without permission)} \tabn
    \hhline{~----}
    \tT & \thrDeDup{Same as (IF-1.a) \rarr \tdupT}{} \tabn
    \hhline{~----}
    \tR & \thrDe{SP denies having sent requests to QTSP on behalf of user}{(or requests without permission)}
        & \prH{4}{1}{4!} & \miNI{User gets Authentication Process request \lbr\Rarr However, lacking \enquote{Blockhain-like} Logging} & \rqMu \tabn
    \hhline{~----}
    \tI & \thrDe{Observing that User requests a signature via SP}{(or that SP requests at QTSP)}
        & \prL{2}{2}{4} & \miOS{Usually, trusted/known networks and TLS} & \rqBa \tabn
    \hhline{~----}
    \tD & \thrDe{User floods SP with requests}{(or SP floods QTSP with requests)}
        & \prL{2}{2}{4} & \miOS{N/A} & \rqBa \tabn
    \hhline{~----}
    \tE & \thrDeDup{Same as (IF-1.b) \rarr \tdupS and \tdupR}{} \tabn
    \addlinespace[-0.19em] \bottomrule
  \end{tabularx}
\end{table}

\newpage
\begin{table}[htb]
  \caption{Full threat-analysis matrix for the Signing Process (SignP) \lbr\dfd components.}
	\label{tab-full-threat-matrix-SignP}
	\centering
  \footnotesize
  \begin{tabularx}{\textwidth}{HDAMR}
    \scriptsize \emph{ST}\cSearch\linebreak\emph{RI}\cSearch\linebreak\emph{DE}\cSearch 
        & \scriptsize \mr{3}{*}{\emph{Threat Description}} 
        & \scriptsize \faArrowAltCircleRight\linebreak\emph{Priority}\linebreak Imp.\emph{\&}Lh.
        & \scriptsize \faShield*\linebreak\emph{Current}\linebreak\emph{MITigation} 
        & \scriptsize \faMountain\linebreak\emph{Requ.}\linebreak\emph{Group}
        \tabularnewline
    \toprule \addlinespace[-0.0em]
    \dfdComp{(SignP-2.a) User signs Challenge-Response} \tabn 
    \addlinespace[-0.19em] \midrule \addlinespace[-0.0em]
    \tS & \thrDe{Other device app uses private binding key}{(signs a different challenge within SE)}
        & \prM{3}{2}{6} & \miOS{Usually, device's SE access through code/biometry input \lbr and link to correct app} & \rqBa \tabn
    \hhline{~----}
    \tT & \thrDe{Signing process modifies storage data}{(e.g., keys)}
        & \prL{3}{1}{3} & \miOS{Device-internal SE implementation and \lbr app's signing program} & \rqBa \tabn
    \hhline{~----}
    \tR & \thrDe{User/App signs challenge while User claiming not to}{}
        & \prH{4}{1}{4!} & \miGE{User gets Authentication Request to access SE's key; 
        hence, a User needs to actively allow this access and should be aware of the access; 
        besides, usually SE elements can be trusted} & \rqAw \tabn
    \hhline{~----}
    \tI & \thrDe{Leakage of user's private binding key}{(other App/Adversary could use key to authenticate a CRA)}
        & \prH{4}{1}{4!} & \miGE{Private key through SE access (usually, well protected) + Adversary would still need User's phone nr. and password to request a QES} & \rqAw \tabn
    \hhline{~----}
    \tD & \thrDe{User/App stops signing challenge (CRA)}{}
        & \prL{3}{1}{3} & \miOS{N/A (might need app reinstall or new User registration)} & \rqBa \tabn
    \hhline{~----}
    \tE & \thrDeDup{Essentially, same as (SignP-2.a) \rarr \tdupS}{} \tabn
    \addlinespace[-0.0em] \toprule \addlinespace[-0.0em]
    \dfdComp{(SignP-3) QTSP signs Document/Hash} \tabn 
    \addlinespace[-0.19em] \midrule \addlinespace[-0.0em]
    \vspace{-1.65em} \tS & \thrDeDup{Essentially, same as (SignP-3) \rarr \tdupE}{\\[-0.45em]\scriptsize(QTSP signs for someone else)} \tabn
    \hhline{~----}
    \tT & \thrDe{Signing process modifies storage data}{(e.g., keys)}
        & \prH{4}{1}{4!} & \miOS{N/A (Handled by QTSP's HSM signing program)} & \rqAw \tabn
    \hhline{~----}
    \vspace{-1.65em} \tR & \thrDeDup{Essentially, same as (SignP-3) \rarr \tdupD and \tdupE}{\\[-0.45em]\scriptsize((i) QTSP signs while claiming not to or (ii) QTSP does not sign while claiming to)} \tabn
    \hhline{~----}
    \tI & \thrDe{Leakage of User's private signing key}{}
        & \prH{4}{1}{4!} & \miNI{Usually, inside HSM \lbr \Rarr Critical part and signing key resides fully at one place remotely} & \rqMu \tabn
    \hhline{~----}
    \tD & \thrDe{\enquote{Signing DoS}}{\lbr(QTSP stops signing QES)}
        & \prH{4}{1}{4!} & \miNI{If HSM failure, QTSP usually has backups; if full QTSP shutdown (not responding at all), no QES can be handed out \Rarr Sole dependency on one QTSP regarding the signing process itself} & \rqMu \tabn
    \hhline{~----}
    \tE & \thrDe{QTSP signs on behalf of User}{(unauthorized Document/Hash signature for the respective User)}
        & \prH{4}{2}{8!} & \miNI{Legal resrictions (on paper/via organizational means) and normally needs User Authentication \Rarr Technically, QTSP can create a QES \lbr(i) without User permission and (ii) without User noticing it} & \rqMu \tabn
    \addlinespace[-0.19em] \bottomrule
  \end{tabularx}
\end{table}

\newpage
\begin{table}[htb]
  \caption{Full threat-analysis matrix for the Data Store (DS) \dfd components.}
	\label{tab-full-threat-matrix-DS}
	\centering
  \footnotesize
  \begin{tabularx}{\textwidth}{HDAMR}
    \scriptsize \emph{ST}\cSearch\linebreak\emph{RI}\cSearch\linebreak\emph{DE}\cSearch 
        & \scriptsize \mr{3}{*}{\emph{Threat Description}} 
        & \scriptsize \faArrowAltCircleRight\linebreak\emph{Priority}\linebreak Imp.\emph{\&}Lh.
        & \scriptsize \faShield*\linebreak\emph{Current}\linebreak\emph{MITigation} 
        & \scriptsize \faMountain\linebreak\emph{Requ.}\linebreak\emph{Group}
        \tabularnewline
    \toprule \addlinespace[-0.0em]
    \dfdComp{(DS-User\rarr SE) Private Binding-Key Store; locally on User's device} \tabn 
    \addlinespace[-0.19em] \midrule \addlinespace[-0.0em]
    \vspace{-1.65em} \tS & \thrDeDup{Essentially, same as (SignP-2.a) \rarr \tdupS}{\\[-0.45em]\scriptsize(we need to trust the SE's implementation to a certain degree)} \tabn
    \hhline{~----}
    \vspace{-1.65em} \tT & \thrDeDup{Essentially, same as (SignP-2.a) \rarr \tdupT}{\\[-0.45em]\scriptsize(we need to trust the SE's implementation to a certain degree)} \tabn
    \hhline{~----}
    \vspace{-1.65em} \tR & \thrDeDup{Essentially, same as (SignP-2.a) \rarr \tdupR}{\\[-0.45em]\scriptsize(we need to trust the SE's implementation to a certain degree)} \tabn
    \hhline{~----}
    \vspace{-1.65em} \tI & \thrDeDup{Essentially, same as (SignP-2.a) \rarr \tdupI}{\\[-0.45em]\scriptsize(we need to trust the SE's implementation to a certain degree)} \tabn
    \hhline{~----}
    \tD & \thrDe{SE/App stops signing}{\lbr(or binding key gets lost/erased)}
        & \prL{3}{1}{3} & \miOS{N/A (Might need app reinstall or new User authentication \lbr\Rarr ideally, User-key backups)} & \rqBa \tabn
    \hhline{~----}
    \vspace{-1.65em} \tE & \thrDeDup{Essentially, same as (SignP-2.a) \rarr \tdupS}{\\[-0.45em]\scriptsize(we need to trust the SE's implementation to a certain degree)} \tabn
    \toprule \addlinespace[-0.0em]
    \dfdComp{(DS-QTSP\rarr HSM) Private Signing-Keys Store; remotely in QTSP's HSM} \tabn 
    \addlinespace[-0.19em] \midrule \addlinespace[-0.0em]
    \vspace{-1.65em} \tS & \thrDeDup{Essentially, same as (SignP-3) \rarr \tdupE}{\\[-0.45em]\scriptsize(QTSP's HSM signs for someone else)} \tabn
    \hhline{~----}
    \tT & \thrDe{Private signing keys are modified}{(leads to incorrect signatures or \enquote{Signing DoS})}
        & \prH{4}{1}{4!} & \miNI{Generally, N/A (handled by QTSP's HSM) \Rarr similarity to (DS-QTSP\rarr HSM) \rarr \tI} & \rqMu \tabn
    \hhline{~----}
    \vspace{-2.55em}\tR & \thrDeDup{Essentially, same as (SignP-3) \rarr \tdupD and \tdupE}{\\[-0.45em]\scriptsize((i) QTSP's HSM signs while claiming not to \\[-0.25em]\scriptsize or (ii) QTSP's HSM does not sign while claiming to)} \tabn
    \hhline{~----}
    \vspace{-1.65em} \tI & \thrDeDup{Essentially, same as (SignP-3) \rarr \tdupI}{\\[-0.45em]\scriptsize(User's private signing key leaks)} \tabn
    \hhline{~----}
    \vspace{-1.65em} \tD & \thrDeDup{Essentially, same as (SignP-3) \rarr \tdupD}{\\[-0.45em]\scriptsize(\enquote{Signing DoS} \rarr QTSP's HSM stops signing QES)} \tabn
    \hhline{~----}
    \vspace{-1.65em} \tE & \thrDeDup{Essentially, same as (SignP-3) \rarr \tdupE}{\\[-0.45em]\scriptsize(QTSP's HSM signs on behalf of User)} \tabn
    \toprule \addlinespace[-0.0em]
    \dfdComp{(DS-SP\rarr QES*) Signature Store; remotely at Service Provider (SP)} \tabn 
    \toprule \addlinespace[-0.0em]
    \tS & \thrDeDup{N/A}{} \tabn
    \hhline{~----}
    \tT & \thrDe{User signatures are modified}{(would invalidate signature and needs re-signing, or signature forgery to another content)}
        & \prL{2}{2}{4} & \miNI{With proper Hash and signing functions, \emph{currently} resilient against adapting signature to other document/data \lbr \Rarr Potential PQ threat} & \rqAw \tabn
    \hhline{~----}
    \tR & \thrDe{SP denies having User QESs'}{}
        & \prL{2}{1}{2} & \miOS{N/A (this would contradicts with SP's business interest of providing Users' a good signing service)} & \rqBa \tabn
    \hhline{~----}
    \tI & \thrDe{User QES leaks}{(respective document could leak too; \lbr hence, probably sensitive data)}
        & \prM{3}{2}{6} & \miOS{N/A (SP needs to properly protect its storage)} & \rqBa \tabn
    \hhline{~----}
    \tD & \thrDe{SP's signature store not reachable or data gets lost}{}
        & \prL{2}{2}{4} & \miOS{N/A (SP needs to properly protect and backup its storage)} & \rqBa \tabn
    \hhline{~----}
    \vspace{-1.65em} \tE & \thrDeDup{Essentially, same as (DS-SP\rarr QES*) \rarr \tdupI}{\\[-0.45em]\scriptsize(SP leaks QES and respective document)} \tabn
    \addlinespace[-0.19em] \bottomrule
  \end{tabularx}
\end{table}
} 







\ifnum\EDIT=1
  \zlabel{endcount}
\fi




\end{document}
